\documentclass[journal=aamick,manuscript=article]{achemso}

\usepackage[version=3]{mhchem} 
\usepackage[T1]{fontenc}       
\usepackage{graphicx}
\usepackage{caption}
\usepackage{subcaption}
\usepackage[flushleft]{threeparttable}

\author{Danila S. Saranin}
\affiliation[NUST MISiS]
{National University of Science and Technology MISiS, Moscow 119049, Russia}
\altaffiliation{Contributed equally to this work}
\author{Abolfazl Mahmoodpoor}
\affiliation[ITMO University]
{ITMO University, Kronverkskiy pr. 49, 197101 St. Petersburg, Russia}
\altaffiliation{Contributed equally to this work}
\author{Pavel M. Voroshilov}
\email{p.voroshilov@metalab.ifmo.ru}
\affiliation[ITMO University]
{ITMO University, Kronverkskiy pr. 49, 197101 St. Petersburg, Russia}
\author{Constantin R. Simovski}
\affiliation[ITMO University]
{ITMO University, Kronverkskiy pr. 49, 197101 St. Petersburg, Russia}
\alsoaffiliation[Aalto University]
{Aalto University, School of Electrical Engineering, Department of Electronics and Nanoengineering, P.O. Box 15500, 00076 Aalto, Finland}
\author{Anvar A. Zakhidov}
\affiliation[ITMO University]
{ITMO University, Kronverkskiy pr. 49, 197101 St. Petersburg, Russia}
\alsoaffiliation[The University of Texas at Dallas]
{The University of Texas at Dallas, Physics Department and The NanoTech Institute, Richardson 75080, USA}

\title{Ionically Gated Small Molecule OPV: Interfacial doping of Charge collector and Transport layer}


\begin{document}

\begin{abstract}
We demonstrate an improvement in the performance of organic photovoltaic (OPV) systems based on small molecules by ionic gating via controlled reversible n-doping of multi-wall carbon nanotube (MWCNT) coated on fullerenes ETL: \ce{C60} and \ce{C70}. Such electric double layer charging (EDLC) doping, achieved by ionic liquid (IL) charging, allows tuning the electronic concentration in MWCNT and in the fullerene planar acceptor layers, increasing it by orders of magnitude. This leads to decreasing the series and increasing the shunt resistances of OPV and allows to use of thick (up to 200 nm) ETLs, increasing the durability and stability of OPV.
Two stages of OPV enhancement are described, upon the increase of gating bias V$\textsubscript{g}$: at small (or even zero) V$\textsubscript{g}$ the extended interface of IL and porous transparent MWCNT  is charged by gating, and the fullerene charge collector is significantly improved, becoming an ohmic contact. This changes the S-shaped I-V curve via improving the electron collection by n-doped MWCNT cathode with ohmic interfacial contact. The I-V curves further improve at higher gating bias V$\textsubscript{g}$ due to the raising of the Fermi level and lowering of MWCNT work function. At the next qualitative stage, the acceptor fullerene layer becomes n-doped by electron injection from MWCNT while ions of IL penetrate into fullerene. At this step the internal built-in field is created within OPV, which helps exciton dissociation and charge separation/transport, increasing further the J$\textsubscript{sc}$ and the FF (Fill factor). Overall power conversion efficiency (PCE) increases nearly 50 times in CuPc/fullerene OPV with MWCNT cathode. The concept of ionically gated MWCNT-ETL interface is numerically simulated by the drift-diffusion model which allows to fit the observed I-V curves. The comparison with ionically gated polymeric bulk heterojunction (BHJ)-OPV with MWCNT-planar ETL interface is presented, showing quite different physical behavior. Demonstrated here ionic gating concept for most simple classical planar small-molecule OPV cells can be potentially applied to more complex highly-efficient hybrid devices, such as perovskite PV with ETL/HTL, providing a new way to tune their properties via controllable and reversible interfacial doping of charge collectors and transport layers.
\end{abstract}

\section{Introduction}

Organic photovoltaics is one of the competitive technologies in the modern renewable energy sector, which is capable to partially meet the needs of power generation due to many advantages such as low cost, extreme flexibility, light weight, and large-area manufacturing \cite{riede2008small}. Since the pioneering work by Ching W. Tang \cite{tang1986two} on two-layer OPV cell based on the molecular donor-acceptor structure of CuPc and perylene tetracarboxylic derivative with a PCE of about 1\%, the significant progress has been achieved in understanding and improvement of OPV systems, leading to a PCE of over 14\% \cite{zhou2019all} in single-junction and over 15\% \cite{ryu2020strategic} in tandem OPV small molecule-based devices that are not so far from the commercialization threshold.

A rather quick increase in the performance of OPV solar cells has happened mainly due to the synthesis of better organic materials with higher charge mobility \cite{sun2005organic, bredas2009molecular}. However, it is not the only way, and further improvement is still possible since those OPV cells are usually undoped bulk heterojunction (BHJ) type structures. If the doping of transport layers can be easily achieved then additional enhancement of efficiency will be a straightforward success. It is well known that p-i-n OPV solar cells have better performance and are more stable due to the thicker p-doped hole transport layer (HTL) and the n-doped electron transport layer (ETL) with low series resistances. In a series of papers \cite{fujii1996organic, drechsel2004high, maennig2004organic, zakhidov1997strategies, yoshino1997novel, siebert2014electroabsorption}, p- and n- type electronic doping of organic donor (D) and acceptor (A) transport layers has been shown to increase the performance of small molecule OPV cells, and 8-10\% efficiency has been achieved in tandems, demonstrating a great promise of p-i-n organic structures \cite{meerheim2014highly}. 

One of the outcomes of creating true  p-i-n organic structures with doped ETL and HTL layers \cite{meerheim2014highly, lussem2013doping, li2006acridine} as opposed to intrinsic, undoped D-A structures (which are commonly and mistakenly called sometimes as p-n diodes), is the advantage of using thick transport layers with very low series resistances namely due to doping of ETL and HTL. So \ce{C60} layers of n-doped A(n) ETL as thick as 100 nm\cite{siebert2014electroabsorption, meerheim2014highly} have been used, as opposed to usual thickness of intrinsic A(i) layer of 7-10 nm. Such thicker layers allow achieving more durable OPV structures without pin-holes. We do not discuss here the obvious advantages of p-i-n structures, such as ohmic contacts with electrodes (and non-sensitivity to work function of an electrode), better charge separation by built-in electric fields, and many others (well described, e.g., in reviews of Karl Leo team \cite{riede2008small}).

Indeed, the truly doped p-i-n planar geometry of small molecule OPV improves the separation of positive and negative charges by built-in potentials, and it also decreases series resistance, enhancing PCE. The most significant progress here is obtained by charge-transfer (CT) type doping the donor layers (e.g. of phthalocyanine by strong acceptor molecules, such as \ce{F4-TCNQ} \cite{lussem2013doping}). Such CT-doping as opposed to substitutional doping (as P and B atoms substitution in Si crystal and other inorganic semiconductors) is usually achieved by intercalation of strong CT dopant between molecules of a host: so most recently and successfully obtained by co-evaporation of strong organic molecular donors (such as acrylic orange, or usual Li atoms) into \ce{C60} layers \cite{li2006acridine}). However, such CT doping is usually done in low molecular OPV systems using air-sensitive dopants in a high vacuum process \cite{drechsel2004high, zakhidov1997strategies, yoshino1997novel}, which is very sophisticated and expensive, and cannot be used for liquid-based air processing of small molecule OPV cells, while the recent progress with PCE of $\sim$ 10\% is namely due to solution processing\cite{Zhou2016Enhancement,Kim2017Performance}.

Doping with the optimum concentration of electrons or holes can modify the physical properties of both A and D layers in organic electronic devices, particularly in OPV and OLED. Therefore, improved ways to achieve carrier doping have been pursued extensively in organic electronics arena, with metal-intercalation, (as mentioned above Li-intercalation into \ce{C60} films) is one of the most important techniques for electron doping of organic/inorganic solids, and has produced not only efficient p-i-n OPV but also \ce{C60} and FeSe superconductors from insulators and metallic solids \cite{takahei2016new}. The most successful examples here are \ce{A_xC60} (where A is Alkali metal) and \ce{C60} superconductors. Such metal or organic molecule (TCNQ, etc.) intercalation has been performed using not only vacuum co-evaporation but also by liquid solvent techniques. Strong donor (e.g., Li, Na, or acrylic orange) intercalation can donate electrons to acceptor ETL layer, which shifts the Fermi level upward.

Recently, the electric-double-layer charging (EDLC) has attracted significant attention as a new way to control the carrier density at the interface of nanoscale materials\cite{Zhao2018Understanding, Lieb2018Ionic}, particularly at the contacts with carbon nanotubes (CNTs) \cite{emmenegger2003investigation} and graphene \cite{uesugi2013electric} which can be achieved reversibly and with no change of chemical composition or structure. Various novel physical properties such as superconductivity, \cite{ueno2008electric, ye2010liquid, bollinger2011superconductor} metal-insulator \cite{yuan2009high} and ferromagnetism \cite{yamada2011electrically} have been shown. We have demonstrated first that tunable polymeric OPV with bulk heterojunction structure can be created using ionic liquid for gating type EDL charging of CNT by the formation of double layer doping of single-wall carbon nanotube (SWCNT) and MWCNT \cite{cook2014ambient, cook2013electrochemically}.

The cathode material is another factor that hinders the performance improvement of OPV cells mostly due to the use of low work function alkali metals like Cs or Li coated cathodes with low air stability. Therefore, Al or Ag electrodes are less attractive for use in OPV cells. Air-stable CNTs have successfully demonstrated their ability to function as a transparent electrode in small molecule OPV cell with doped HTL and ETL layers \cite{kim2012semi}. However, such n-i-p OPV device with n-doped \ce{C60} and CNT anode on the top of p-type HTL have a certain drawback since it was prepared in a high vacuum by co-evaporation doping of \ce{C60} with a very expensive dopant. This inspired us to develop a new architecture deprived of this shortcoming. Here, we propose another structure that is more advantageous since n-doping of both MWCNTs and fullerene can be done in ambient conditions with no need for any additional high-priced processing. MWCNTs have initially high sheet resistance R$\textsubscript{sheet}$ exceeding 1000 $\Omega/sq$ in the undoped as-synthesized state that is very high for PV applications. Usually, high work function of MWCNTs in our OPV cell can be substantially modified by EDLC under IL gating similar to our earlier \cite{cook2014ambient, cook2013electrochemically} for BHJ poly-OPV, turning them into a good cathode. In other words, the Fermi level of MWCNTs can be raised up by n-type doping, thereby allowing better electron collection from the active layer of OPV cell, and the value of R$\textsubscript{sheet}$ can be decreased to the acceptable level below 100 $\Omega/sq$. In an EDLC ionic gating method, the carriers can first accumulate around the extended interface of highly porous nanomaterial, such as CNT, and their concentration and work function can be controlled by an electric field of the gate V$\textsubscript{g}$. Such simple reversible and tunable EDLC-doping has never been previously used in small molecule OPV cells.

So the motivation of our present paper is to create a small molecule OPV device in which n-doping can be easily achieved in a system of porous CNT electrode coated on top of fullerene film ETL of a most simple two-layer OPV: CuPc/fullerene with methods of reversible ionic EDLC. This requires new architecture, which combines IL supercapacitor, connected in parallel with OPV as we introduced earlier for a polymeric OPV with IL \cite{cook2013electrochemically}, and demonstrated its advantages in tandems \cite{saranin2017tunable}.

In this paper, we apply the gating in IL to classical small molecule OPV with \ce{DEME-BF4} IL and study the dynamics of this ionically gated system of MWCNT@\ce{Fullerene} (i.e. cathode@ETL) to understand how it accumulates electrons at two interfaces: the extended interface of CNT with IL by doping the CNT by n-type first, and then at the interface of CNT with \ce{C60} ETL by further propagating the n-doping into the ETL, and how the ions move first in MWCNT extended interfaces network by creating EDLC charging at the atomic layer of tubes, and then into and through fullerene molecular ETL layer of different thickness creating n-doping in \ce{C60} by electrochemical electronic charge injection from CNT that electrically neutralizes the positive ions of ionic liquid. We show here that this process differs significantly from the case of polymeric OPV, studied in our earlier paper \cite{cook2013electrochemically}, since in a thick molecular fullerene film (with molecules bonded by Van der Waals forces) there are clear three stages of charging and n-doping of separately MWCNT and fullerene subsystems. To compare the difference of ionic gating in planar small molecule OPV and BHJ poly-OPV we shortly compare the dynamics of the I-V curves in both systems upon gating voltage increase. And then we describe the microscopic modeling by drift-diffusion computer simulations the gating and by fitting the observed parameters, discuss the physical processes at interfaces upon ions diffusion.

\section{Methods}

Small molecule OPV cell was prepared in the multi-source resistive glove box integrated high-vacuum system (Angstrom Engineering Inc., Canada) by the sequential thermal evaporation onto UV-ozone treated, patterned ITO-glass substrates of the following layers: a 7 nm thick copper(II) phthalocyanine (from H.W. Sanders  Corp) HTL, a 60 nm thick co-evaporated with a smooth gradient CuPc:fullerene mixed D:A layer and top fullerene layer of variable thickness. The following fullerene materials were used: \ce{C60} or \ce{C70} (both >98\% from Nano-C) for different devices. The base pressure of the chamber was kept around 10$^{-8}$ $mbar$ during the evaporation.

\begin{figure}[htbp]
\centering
\includegraphics[width=1\textwidth]{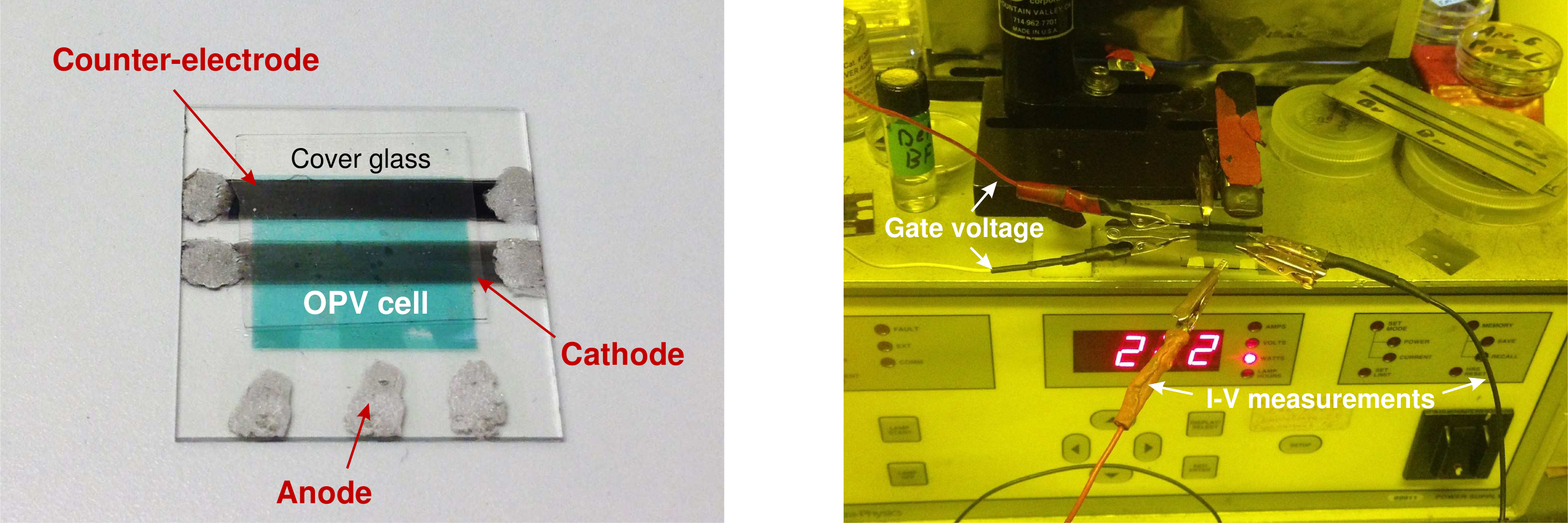}
\caption{View of OPV device with laminated semi-transparent MWCNT cathode and MWCNT counter-electrode soaked by ionic liquid (left) and measurement setup and electrical connection scheme for determination of solar cell parameters when gate voltage V$\textsubscript{g}$ applied (right).}
\label{photo}
\end{figure}

MWCNT forest was produced by a chemical vapor deposition (CVD) process. Our OPV devices contain three electrodes: the anode of the solar cell (ITO under CuPc), the cathode (first MWCNT sheet on top of fullerene ETL), and the counter-electrode (second MWCNT sheet). Both MWCNT cathode and another MWCNT electrode named "counter-electrode" or gate were deposited by manual dry lamination on the top of the organic multilayer structure outside the glovebox in ambient conditions. After that, the deposited MWCNTs were immersed in liquid hydrofluoroether (HFE) solvent for several seconds to condense tube to tube interconnects and therefore improve conductivity and stability of the electrodes. Contacts for cathode, counter-electrode, and ITO-anode were created using silver paint. A small amount of ionic liquid, $N,N$-Diethyl-$N$-methyl-$N$-(2-methoxyethyl) ammonium tetrafluoroborate, \ce{DEME-BF4} (Kanto Chemical Co. Inc.), was dropped on top of both MWCNT electrodes. A thin transparent glass cover-slip was placed over the ionic liquid. A general view of one fabricated OPV device is shown on the left part of Fig. \ref{photo}.

\begin{figure}[htbp]
\centering
\includegraphics[width=0.8\textwidth]{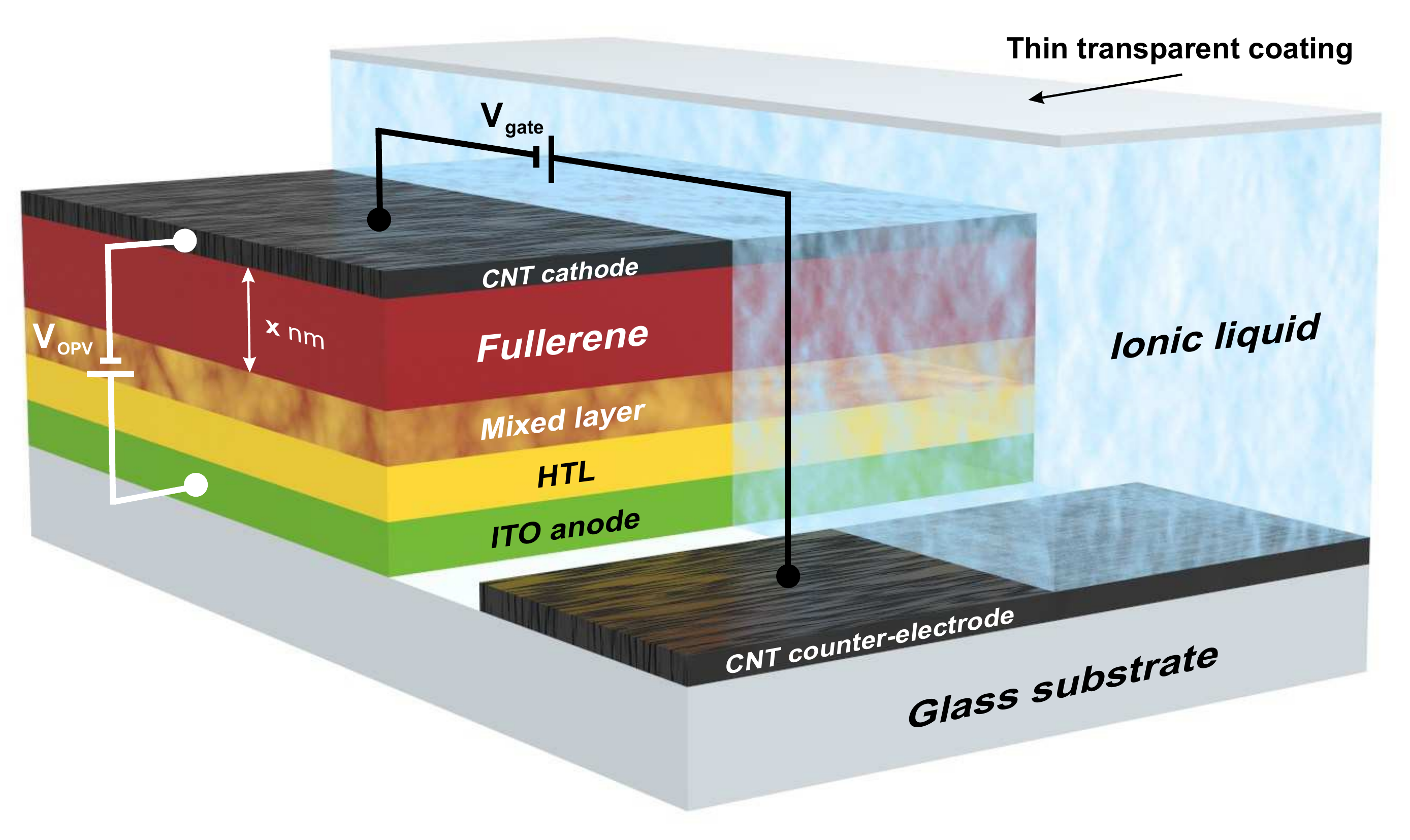}
\caption{Side view of our OPV-CNT-IL test cell. The light shines through the bottom transparent conductive oxide cathode (ITO) and into the CuPc/fullerene mixed active layer. Ions can EDLC charge on the surface of CNT and move further through the porous CNT into the small molecule fullerene matrix of OPV, doping adjacent CNT cathode layers which are tested for different thicknesses up to 200 nm.}
\label{structure}
\end{figure}

OPV devices were characterized with an AM 1.5G solar simulator calibrated to one sun (100 $mW/cm^2$) and two LabVIEW controlled Keithley 2400 source measure units (SMU) in a nitrogen glove box. One of these SMUs was used to apply gate voltage (V$\textsubscript{g}$) by connecting counter-electrode with MWCNT cathode while the second SMU measured I-V parameters of OPV cells through connection to the anode and cathode of the OPV part (see right part of Fig. \ref{photo}). Application of bias V$\textsubscript{g}$ between the MWCNT cathode and MWCNT counter-electrode would thus produce super-capacitive EDLC while measurements under of $I$ and $V$ between the ITO anode and MWCNT cathode would result in the photovoltaic I-V curve from which we get V$\textsubscript{oc}$ and J$\textsubscript{sc}$ response. Such EDLC creates an asymmetry between the anode (ITO) and cathode by the decreased work function of the MWCNTs forming a built-in electric field that allows better charge collection.

\section{Results and discussions}

\subsection{Experimental results for small molecule OPV}

\begin{figure*}[htbp]
    \centering
    \begin{subfigure}[t]{0.44\textwidth}
        \centering
        \includegraphics[width=1\textwidth]{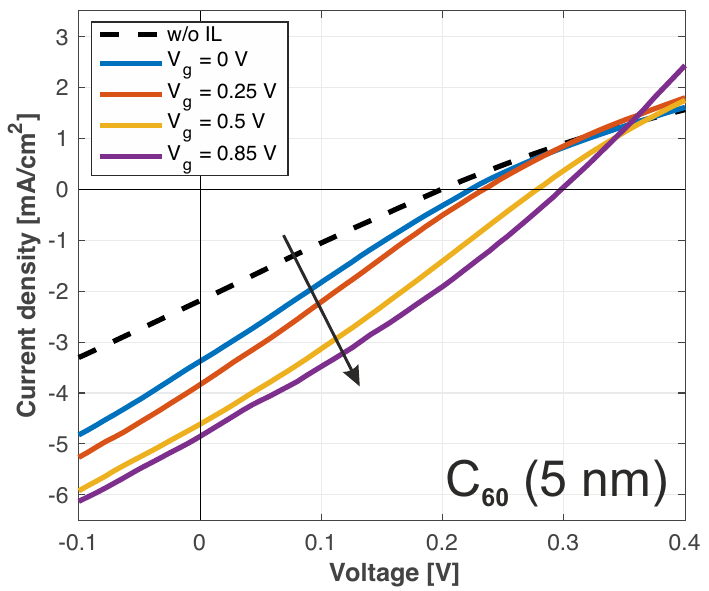}
        \caption{}
    \end{subfigure}%
    ~ 
    \begin{subfigure}[t]{0.455\textwidth}
        \centering
        \includegraphics[width=1\textwidth]{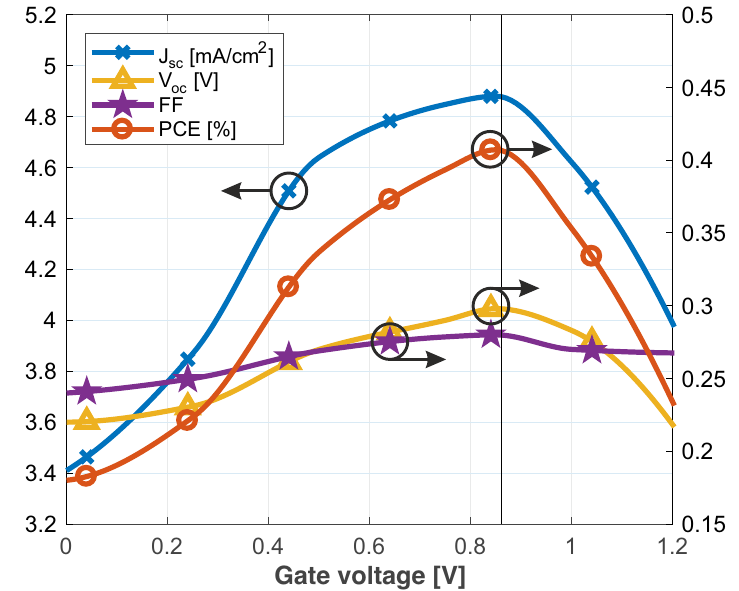}
        \caption{}
    \end{subfigure}
    \caption{I-V curves of the OPV solar cell with a 5 nm-thick \ce{C60} ETL at different values of gate voltage under illumination (a) and the extracted parameters for the same solar cell as a function of gate voltage (b). OPV device reaches its best performance at V$\textsubscript{g}$ = 0.85 $V$.}
    \label{C60_5nm_all}
\end{figure*}

We measured I-V characteristics of several OPV devices differing by fullerene type (\ce{C60} or \ce{C70}) and ETL thickness (from 5 nm to 200 nm) in two regimes: without the IL and with the IL on the top of MWCNTs at various gate voltage applied. All ungated devices showed a very poor output performance for both types of fullerene ETL due to unoptimized architecture. Namely, the initial difference in Fermi levels of the anode and cathode was very small that resulted in low values of V$\textsubscript{oc}$ ($\sim$ 0.25 $V$) for all devices. High contact resistance at the interface of the cathode and ETL led to sufficiently decreased FF. The highest short-circuit current at initial conditions was observed for a device with 5 nm-thick ETL and its value didn't exceed 2.2 $mA/cm^2$ (Fig. \ref{C60_5nm_all}a, black dashed curve) while with increasing the thickness of fullerene layer J$\textsubscript{sc}$ decreased accordingly. 

\begin{figure*}[t!]
\makebox[\linewidth][c]{%
\begin{subfigure}[b]{.5\textwidth}
\centering
\includegraphics[width=.70\textwidth]{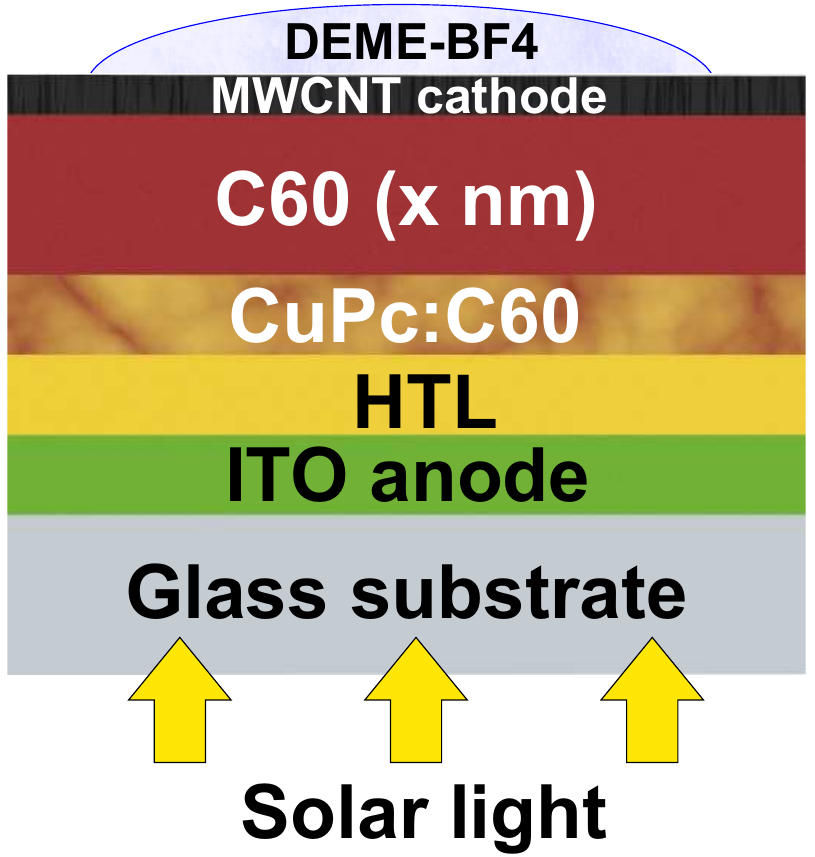}
\caption{}
\end{subfigure}%
\begin{subfigure}[b]{.5\textwidth}
\centering
\includegraphics[width=.9\textwidth]{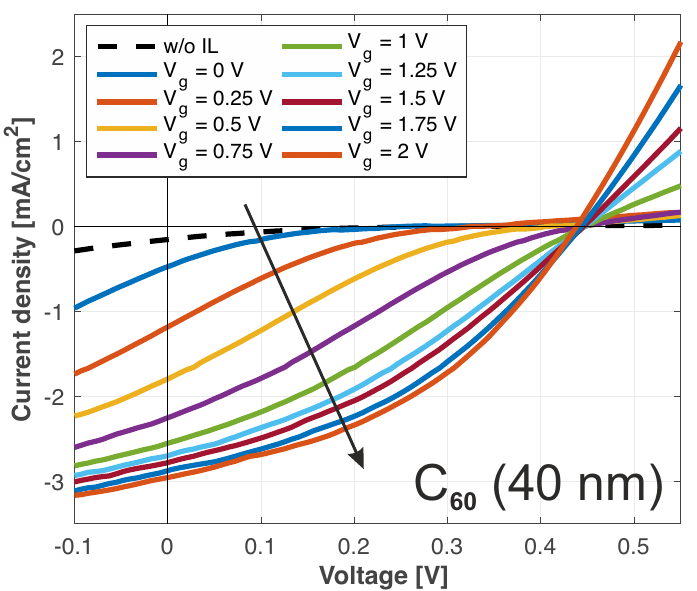}
\caption{}
\end{subfigure}%
}\\
\makebox[\linewidth][c]{%
\begin{subfigure}[b]{.5\textwidth}
\centering
\includegraphics[width=.9\textwidth]{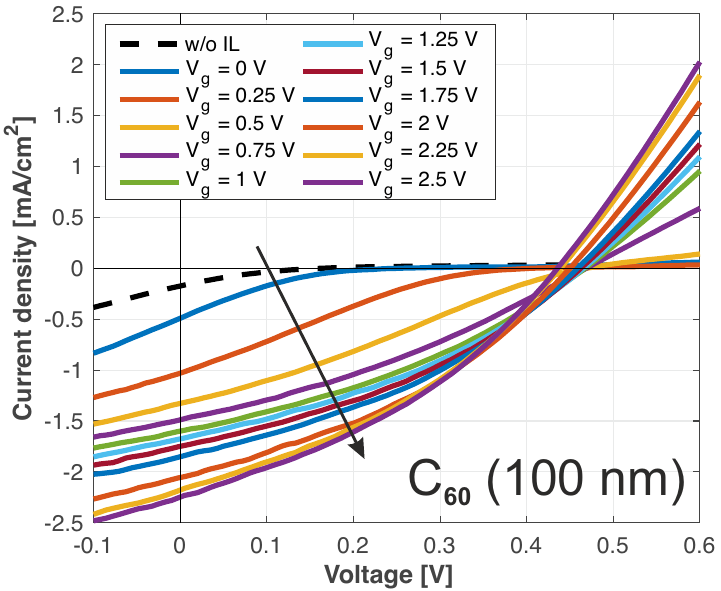}
\caption{}
\end{subfigure}%
\begin{subfigure}[b]{.5\textwidth}
\centering
\includegraphics[width=.9\textwidth]{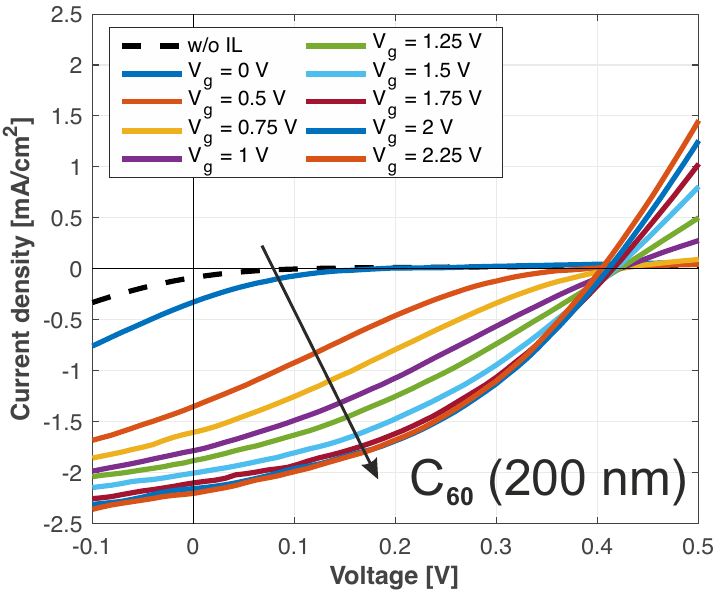}
\caption{}
\end{subfigure}%
}
\caption{(a) Cross-section schematics of the OPV device with top \ce{C60} ETL and laminated MWCNT cathode soaked in \ce{DEME-BF4} IL. (b-d) I-V curves at different values of gate voltage under illumination for 40, 100 and 200 nm thick \ce{C60} ETL.}
\label{С60_sweep_curves}
\end{figure*}

We found that solar cell performance can be improved by the IL even without applying any gate voltage between the counter-electrode and cathode. The most dramatic improvement was observed for J$\textsubscript{sc}$ which jumped to 3.4 $mA/cm^2$ (>50\% growth) for a device with 5 nm-thick \ce{C60} while both FF and V$\textsubscript{oc}$ grew slightly. Similar behavior was shown by other devices with thicker ETL where J$\textsubscript{sc}$ was even doubled due to the IL.

\begin{figure}[htbp]
\centering
\includegraphics[width=.95\textwidth]{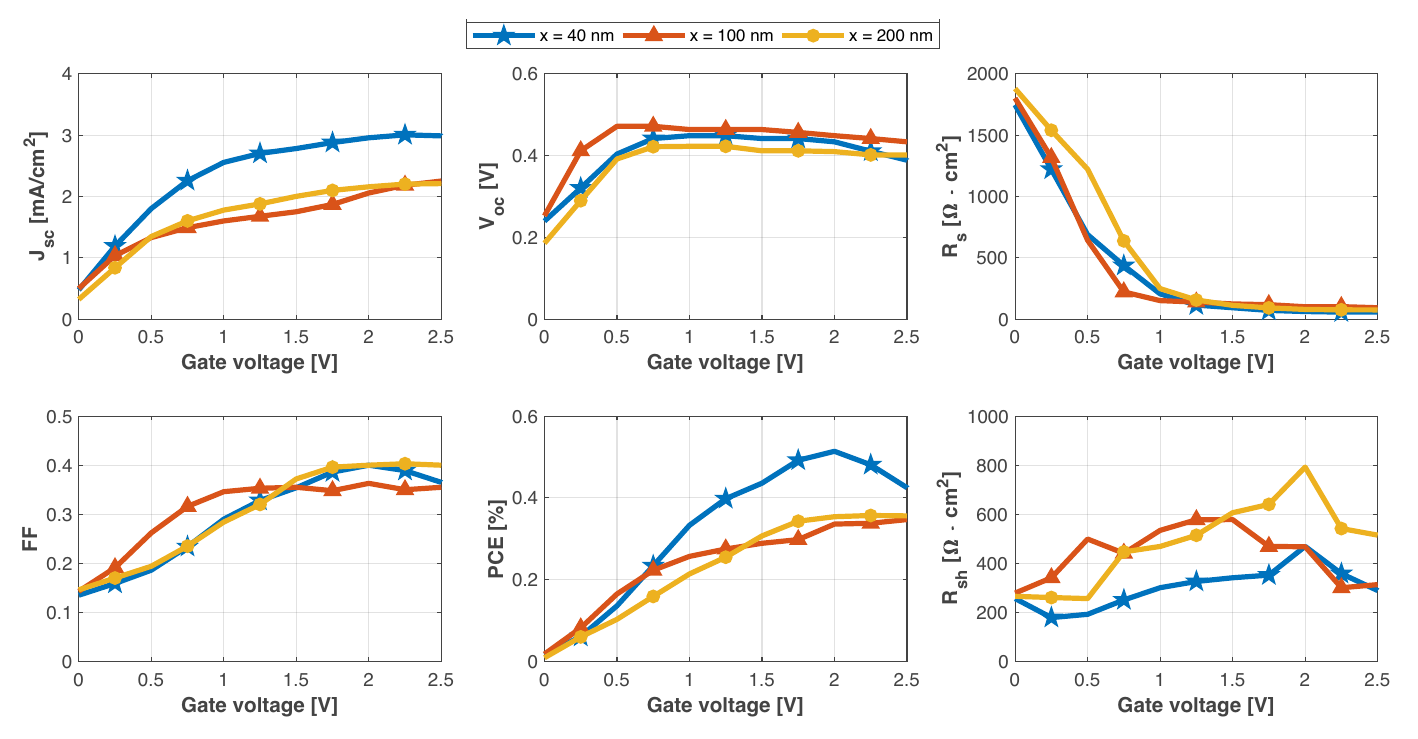}
\caption{Extracted parameters for the OPV cell with \ce{C60} ETL of different thickness as a function of gate voltage.}
\label{C60_sweep_parameters}
\end{figure}

\begin{figure*}[t!]
\makebox[\linewidth][c]{%
\begin{subfigure}[b]{.5\textwidth}
\centering
\includegraphics[width=.70\textwidth]{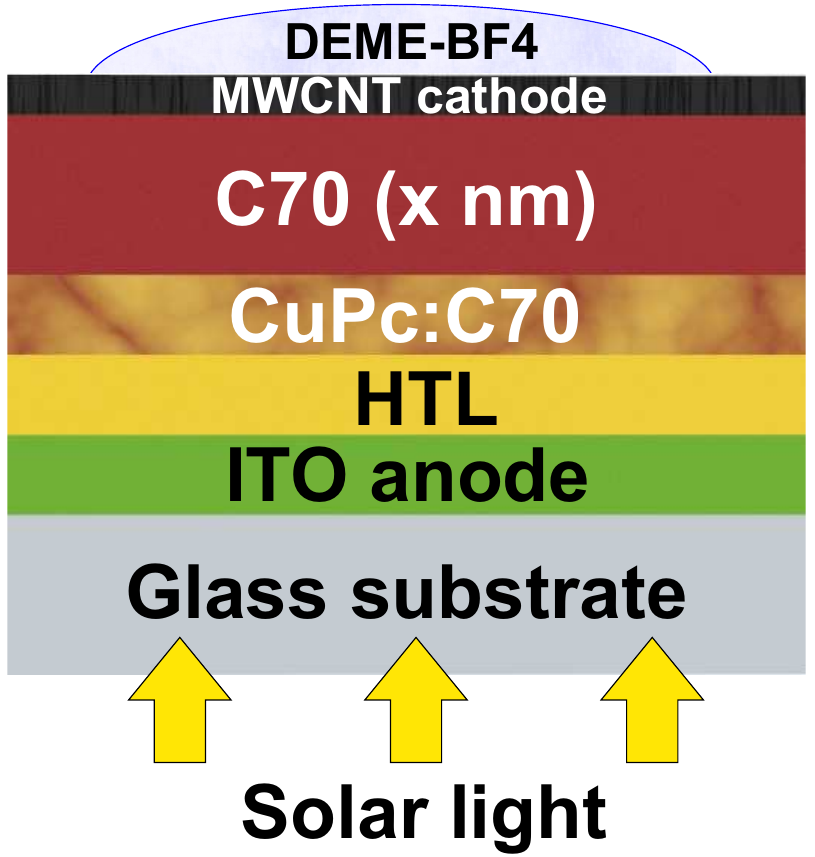}
\caption{}
\end{subfigure}%
\begin{subfigure}[b]{.5\textwidth}
\centering
\includegraphics[width=.9\textwidth]{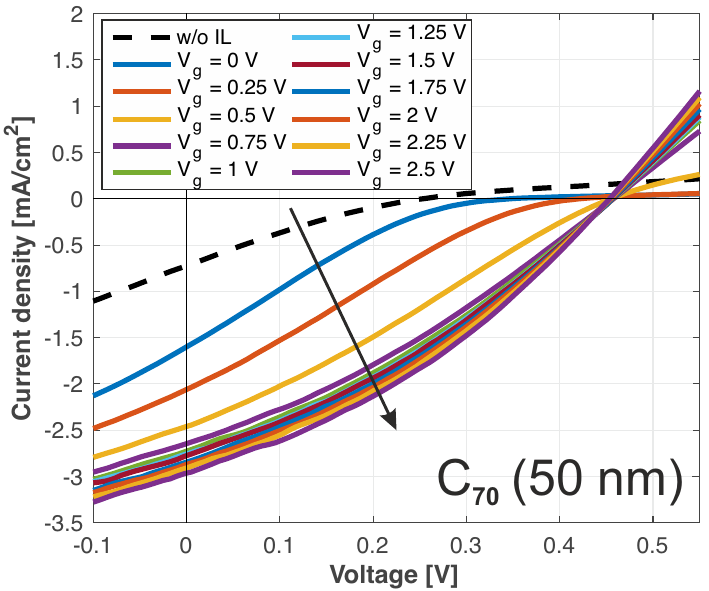}
\caption{}
\end{subfigure}%
}\\
\makebox[\linewidth][c]{%
\begin{subfigure}[b]{.5\textwidth}
\centering
\includegraphics[width=.9\textwidth]{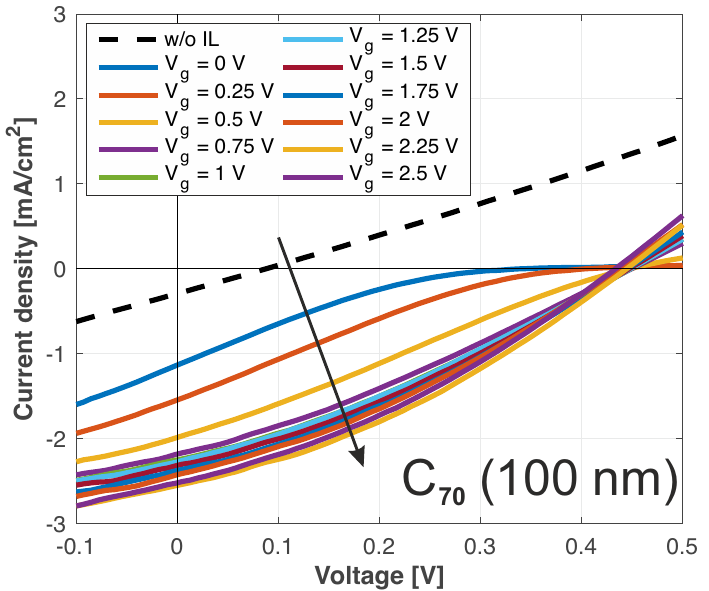}
\caption{}
\end{subfigure}%
\begin{subfigure}[b]{.5\textwidth}
\centering
\includegraphics[width=.9\textwidth]{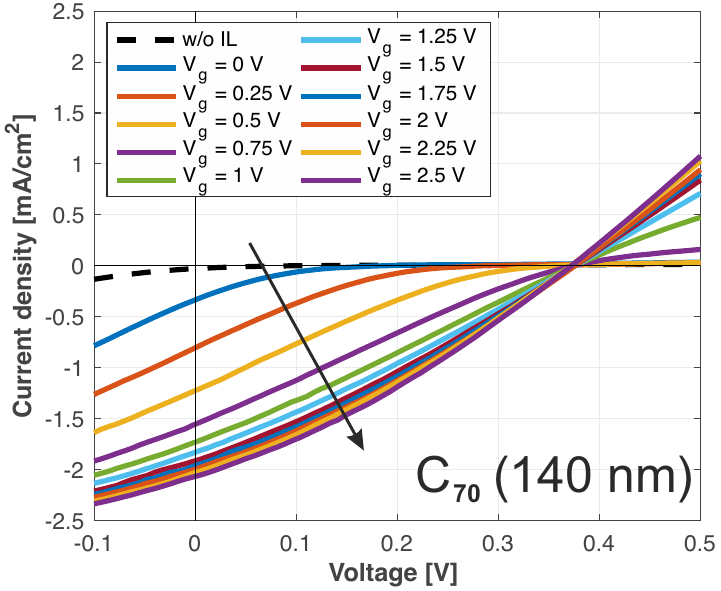}
\caption{}
\end{subfigure}%
}
\caption{(a) Cross-section schematics of the OPV device with top \ce{C70} ETL and laminated MWCNT cathode soaked in \ce{DEME-BF4} IL. (b-d) I-V curves at different values of gate voltage under illumination for 50, 100 and 140 nm thick \ce{C70} ETL.}
\label{C70_sweep_curves}
\end{figure*}

\begin{figure}
\centering
\includegraphics[width=.95\textwidth]{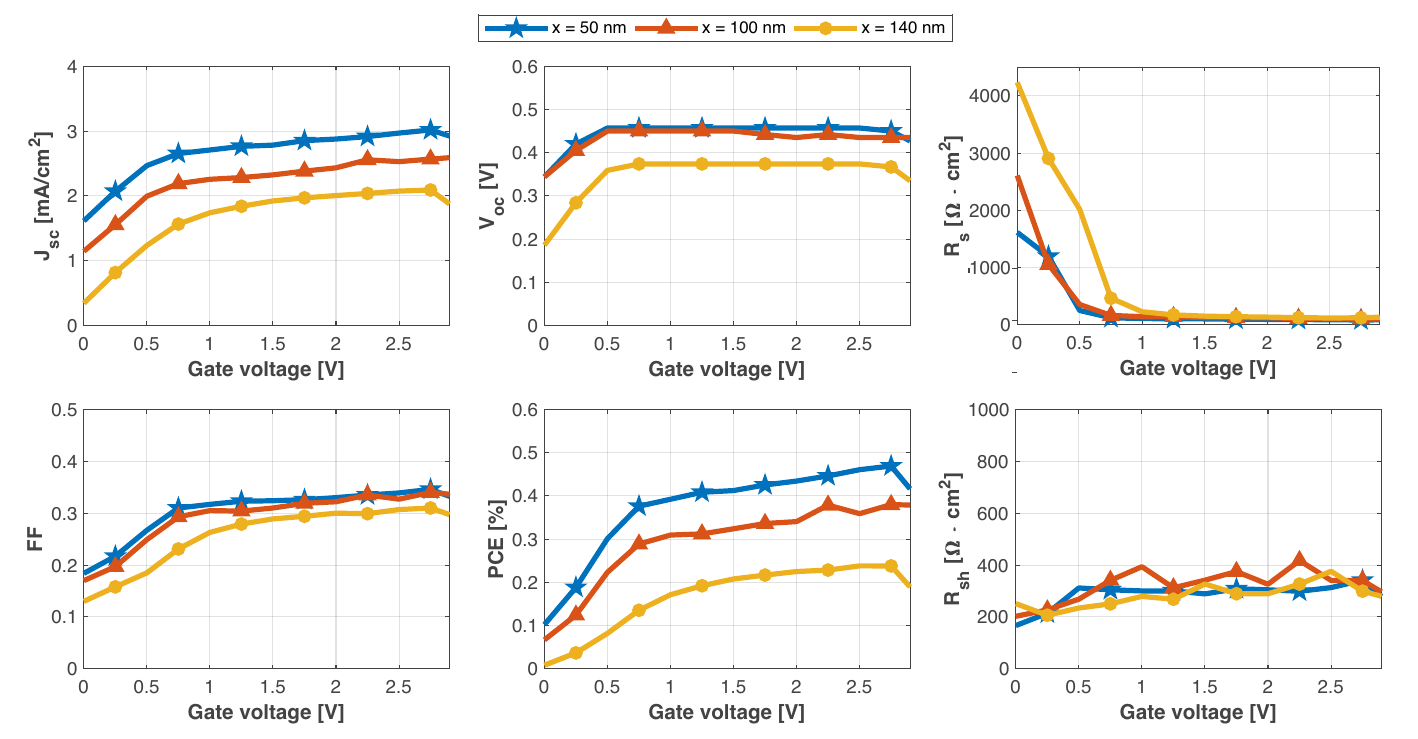}
\caption{Extracted parameters for the OPV solar cell with \ce{C70} ETL of different thickness as a function of gate voltage.}
\label{C70_sweep_parameters}
\end{figure}

Next, we applied different gate voltages between the counter-electrode and the MWCNT@C60 (cathode on ETL) part of our OPV-IL system. In our experiment, the charging time was equal to 2 mins. The dependence of J$\textsubscript{sc}$, V$\textsubscript{oc}$ and FF as a function of the applied gate voltage (V$\textsubscript{g}$) from 0 to 3 $V$ was measured. In Figs. \ref{C60_5nm_all}, \ref{С60_sweep_curves}, \ref{C70_sweep_curves} we show set of I-V curves corresponding to various V$\textsubscript{g}$ values for different thickness and type of fullerene. Extracted OPV parameters plotted in Figs. \ref{C60_sweep_parameters} and \ref{C70_sweep_parameters} are summarized in Tables \ref{tbl:C60_40nm} and \ref{tbl:C70_50nm} for \ce{C60} and \ce{C70}-based devices, respectively. 

\begin{table}
  \caption{Output parameters for the ionically gated OPV cell with 40 nm-thick \ce{C60} ETL.}
  \label{tbl:C60_40nm}
  \begin{tabular}{lllllll}
    \hline
    \hline
    V$\textsubscript{g} (V)$	&	J$\textsubscript{sc} (mA/cm^2)$	&	V$\textsubscript{oc} (V)$	&	FF	&	PCE $(\%)$	&	R$\textsubscript{s} (\Omega \cdot cm^2)$	&	R$\textsubscript{sh} (\Omega \cdot cm^2)$\\
    \hline
	dry & 0.154	&	0.276	&	0.145	&	0.0062	&	...	&	...\\
	0	&	0.479	&	0.240	&	0.135	&	0.016	&	1748	&	256\\
	0.25	&	1.193	&	0.321	&	0.159	&	0.061	&	1222	&	178\\
	0.5	&	1.805	&	0.404	&	0.186	&	0.136	&	690	&	192\\
	0.75	&	2.259	&	0.442	&	0.234	&	0.234	&	437	&	250\\
	1	&	2.558	&	0.449	&	0.290	&	0.333	&	207	&	300\\
	1.25	&	2.704	&	0.449	&	0.328	&	0.398	&	115	&	326\\
	1.5	&	2.784	&	0.442	&	0.354	&	0.436	&	93	&	341\\
	1.75	&	2.881	&	0.442	&	0.386	&	0.492	&	72	&	351\\
	\textbf{2}	&	\textbf{2.959}	&	\textbf{0.434}	&	\textbf{0.400}	&	\textbf{0.514}	&	\textbf{62}	&	\textbf{469}\\
	2.25	&	3.007	&	0.411	&	0.389	&	0.481	&	58	&	357\\
	2.5	&	2.990	&	0.389	&	0.365	&	0.425	&	58	&	289\\
	3	&	1.602	&	0.080	&	0.230	&	0.030	&	53	&	48\\
    \hline
    \hline
  \end{tabular}
\end{table}

OPV device with a 5 nm-thick \ce{C60} layer reaches maximal values of all the solar cell parameters at the certain gate voltage applied (V$\textsubscript{g}$ = 0.85V) and degrades as the increase of V$\textsubscript{g}$ further continues beyond this threshold due to EDLC formation on the CuPc part while devices with thicker ETL are saturated at much higher V$\textsubscript{g}$  > 2 $V$. The level of this threshold is determined by the ETL thickness. The maximum value of J$\textsubscript{sc}$ is 4.9 $mA/cm^2$, however the increased values of FF and V$\textsubscript{oc}$ are still quite modest (0.28 and 0.3 $V$, respectively). For thick devices, we can clearly distinguish the two following stages. At V$\textsubscript{g}$ < 1 $V$ parasitic series resistance R$\textsubscript{s}$ quickly drops by one order while all PV parameters rapidly increase. This improvement is caused by the formation of ohmic contact between MWCNT and fullerene as well as due to the increase in conductivity of the MWCNT electrode. At the next stage (V$\textsubscript{g}$ > 1 $V$), ions start to penetrate into fullerene and it becomes n-type doped increasing further J$\textsubscript{sc}$ and FF. As a result, for \ce{C60}-based (40 nm) device we observe almost 20 times increased J$\textsubscript{sc}$, nearly twice increased both V$\textsubscript{oc}$ and FF that results in significantly increased PCE -- from the value less than 0.01\% to 0.514\% (more than 50 times). A similar trend is demonstrated by other devices.

Lowered J$\textsubscript{sc}$ of our OPV device with MWCNTs in comparison with conventional non-transparent cathodes is caused by the reduced optical absorption in the bulk heterojunction since metals reflect more unabsorbed light than semi-transparent CNTs that was already investigated in details by optical simulations for different transport layer thicknesses in our previous work \cite{kim2012semi}. Another reason for the overall low PCE is the fact that these devices are unoptimized. Since this is a new device one should optimize the OPV thickness and IL volume. We believe that increasing OPV thickness and decreasing IL volume will increase PCE significantly. Small deviations from the trend can be interpreted as variations caused by manual processing and non-uniformity of materials used from different batches. 

Let us consider the physical and photo-electrochemical processes in this IL-OPV device in more detail. In small molecule OPV structure, the photon absorbed in CuPc (D-part) of BHJ creates an exciton, which dissociates at fullerene interface (acceptor part) by electron injection to LUMO of fullerene.  This electron is further collected via the i-n build-in field fullerene(i)-fullerene(n) at the porous MWCNT network cathode of OPV. Positive charges in the CuPc molecules are collected by ITO anode which charges it positively, generating initial V$\textsubscript{oc}$. Positively charged ions from the ionic chamber redistribute around the negatively charged MWCNT cathode, creating EDLC on each nanotube or nanotube bundle and stabilizing the desired n-doping of the cathode. This doping raises the Fermi level in the MWCNT cathode. Moreover, some ions distribute further between the molecules of fullerene films which stabilizes the photogenerated electrons, creating a narrow n-doped layer around CNT and forming a desired ohmic contact at the interface with the MWCNT cathode. Namely this ohmic contact is responsible for eliminating the initial (in dry conditions) S-shaped I-V curve at V$\textsubscript{oc}$ region.

\begin{table}
  \caption{Output parameters for the ionically gated OPV cell with 50 nm-thick \ce{C70} ETL.}
  \label{tbl:C70_50nm}
  \begin{tabular}{lllllll}
    \hline
    \hline
    V$\textsubscript{g} (V)$	&	J$\textsubscript{sc} (mA/cm^2)$	&	V$\textsubscript{oc} (V)$	&	FF	&	PCE $(\%)$	&	R$\textsubscript{s} (\Omega \cdot cm^2)$	&	R$\textsubscript{sh} (\Omega \cdot cm^2)$\\
    \hline
	dry	&	0.733	&	0.254	&	0.199	&	0.037	&	...	&	...\\
	0	&	1.614	&	0.344	&	0.184	&	0.102	&	1616	&	167\\
	0.25	&	2.071	&	0.420	&	0.217	&	0.188	&	1197	&	214\\
	0.5	&	2.467	&	0.457	&	0.267	&	0.301	&	252	&	311\\
	0.75	&	2.653	&	0.457	&	0.310	&	0.376	&	125	&	304\\
	1	&	2.708	&	0.457	&	0.317	&	0.392	&	111	&	300\\
	1.25	&	2.764	&	0.457	&	0.323	&	0.408	&	107	&	300\\
	1.5	&	2.783	&	0.457	&	0.324	&	0.412	&	105	&	288\\
	1.75	&	2.852	&	0.457	&	0.326	&	0.425	&	98	&	308\\
	2	&	2.877	&	0.457	&	0.330	&	0.434	&	96	&	304\\
	2.25	&	2.915	&	0.457	&	0.335	&	0.446	&	94	&	298\\
	2.5	&	2.969	&	0.457	&	0.339	&	0.461	&	90	&	313\\
	\textbf{2.75}	&	\textbf{3.017}	&	\textbf{0.450}	&	\textbf{0.346}	&	\textbf{0.469}	&	\textbf{86}	&	\textbf{341}\\
	3	&	2.861	&	0.411	&	0.323	&	0.381	&	90	&	237\\
    \hline
    \hline
  \end{tabular}
\end{table}

The improvement in the solar cell can be observed with the improvement of the current-voltage characteristics. The initial S-shaped I-V curve with low solar cell parameters improves into one with significantly increased J$\textsubscript{sc}$, FF, and V$\textsubscript{oc}$, as it was mentioned earlier for different fullerene thicknesses. The device can be viewed as an OPV and supercapacitor connected in parallel via a common ion-porous MWCNT cathode. To maintain overall charge neutrality in the electrolyte, the negative ions (\ce{BF4}) move towards a MWCNT counter-electrode that is placed at the opposite side of the IL chamber, forming a second EDLC at the extended interfaces of CNT of counter-electrode, charging it positively and thus creating p-type doping of counter-electrode. The formation of the EDLC on the MWCNT counter-electrode is the charging voltage in the supercapacitor component of the structure. Formation of EDLC is expected to result in the n-doping of both the MWCNT electrode and the fullerene semiconductor, leading to the formation of ohmic contacts and n-i build-in junctions first inside the thick \ce{C60}: n-doping and thus i-n junction propagating through the thickness of planar \ce{C60} and finally reaching into \ce{C60} part of BHJ interpenetrating network. 

within the BHJ of CuPc/fullerene.

\begin{figure*}[htbp]
    \centering
    \begin{subfigure}[t]{0.445\textwidth}
        \centering
        \includegraphics[width=1\textwidth]{fullenes_layer1.pdf}
        \caption{}
    \end{subfigure}%
    ~ 
    \begin{subfigure}[t]{0.445\textwidth}
        \centering
        \includegraphics[width=1\textwidth]{fullenes_layer2_2.pdf}
        \caption{}
    \end{subfigure}
    \caption{Ion penetration from the IL into MWCNTs and fullerene layer at low gate voltage (a) and high gate voltage (b). We show here only i-n in the planar ETL itself, while i-n in the BHJ takes place deeper in the structure, but namely this later i-n improves collection and J$\textsubscript{sc}$ further on.}
    \label{IL_penetration}
\end{figure*}

Summarizing, we have introduced here a concept of n-doped extended CNT interface collector and also a concept of n-doped ETL, i.e. the "thick n-doped ETL based" ionically gated small molecule OPV. An electrolyte or ionic liquid in a microchamber is added on top of the highly porous network of CNT cathode of OPV. Essential for this design is the existence of counter-electrode, which has similarly high (or even higher) capacitance and accommodates negative ions of IL and thus allows positive ions of IL to penetrate further into \ce{C60} ETL, causing its negative n-doping. In our proof of concept studies, we have used MWCNT as such a counter-electrode. In addition to filling the earlier criteria, MWCNT can be easily doped by electrochemical double layer charging, which occurs naturally during the operation of the ionic-OPV structure. Ions from the ionic microchamber provide counter ions for EDLC charging, thus permitting both n-type doping (stabilized by in EDLC by positive ions of DEME) or p-type doping of counter-electrode (stabilized by negative ions in second EDLC). The organic layers adjacent to MWCNT cathode also are n-doped at higher V$\textsubscript{g}$ in the properly built and operated device, and this doping provides not only a better ohmic contact between the doped MWCNT cathode and n-doped fullerene organic layer. This process reconfigures the OPV from undoped D-A type BHJ to n-i-i, or more correctly heterojunction of the: CNT(n)-A(n)-A(i)-D(i)-ITO hybrid (planar with BHJ layer) type OPV with ohmic contacts to electrodes and internal build-in fields, increasing the performance.

\subsection{Comparison of ionic gating dynamics in small molecule OPV with polymeric BHJ-OPV}

\begin{figure*}[ht]
    \centering
    \begin{subfigure}[t]{0.445\textwidth}
        \centering
        \includegraphics[width=1\textwidth]{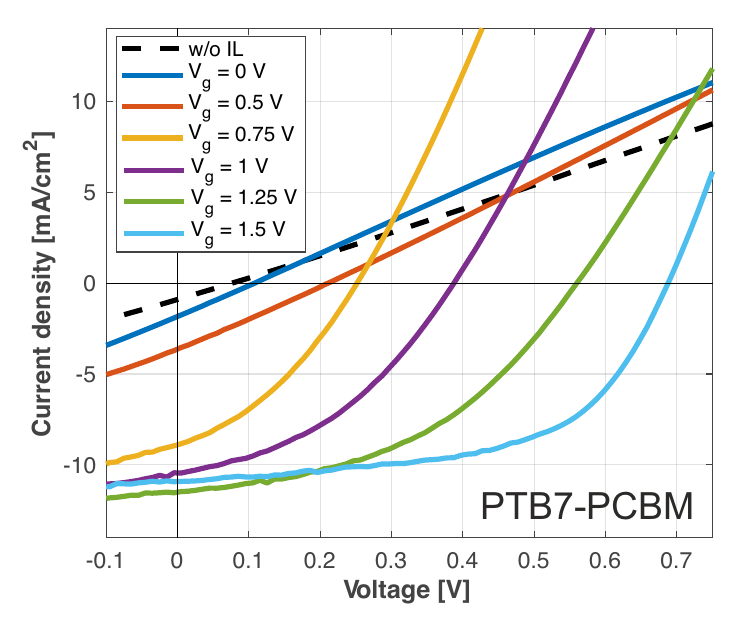}
        \caption{}
    \end{subfigure}%
    ~ 
    \begin{subfigure}[t]{0.445\textwidth}
        \centering
        \includegraphics[width=1\textwidth]{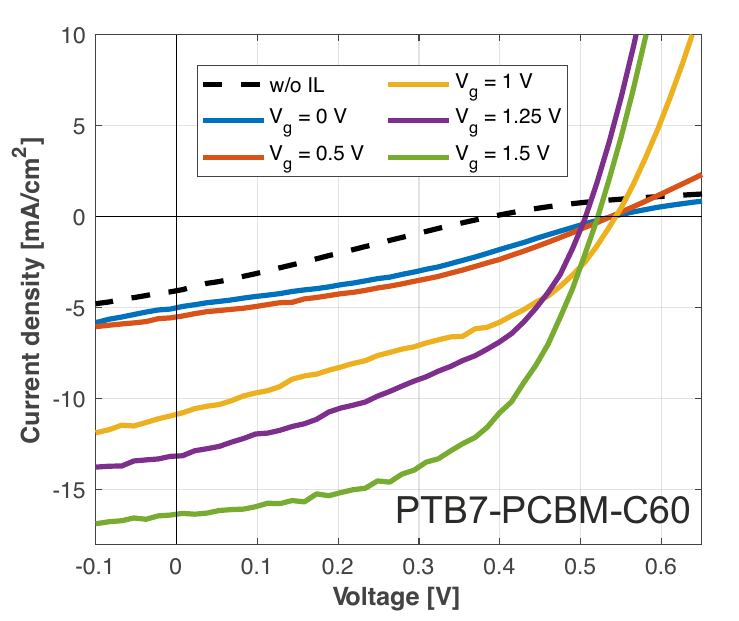}
        \caption{}
    \end{subfigure}
    \caption{I-V curves at different values of gate voltage under illumination for (a) PTB7-PCBM and (b) PTB7-PCBM-\ce{C60} polymeric BHJ-OPV devices.}
    \label{sm_bhj1}
\end{figure*}

\begin{figure*}[ht]
    \centering
    \begin{subfigure}[t]{0.325\textwidth}
        \centering
        \includegraphics[width=1\textwidth]{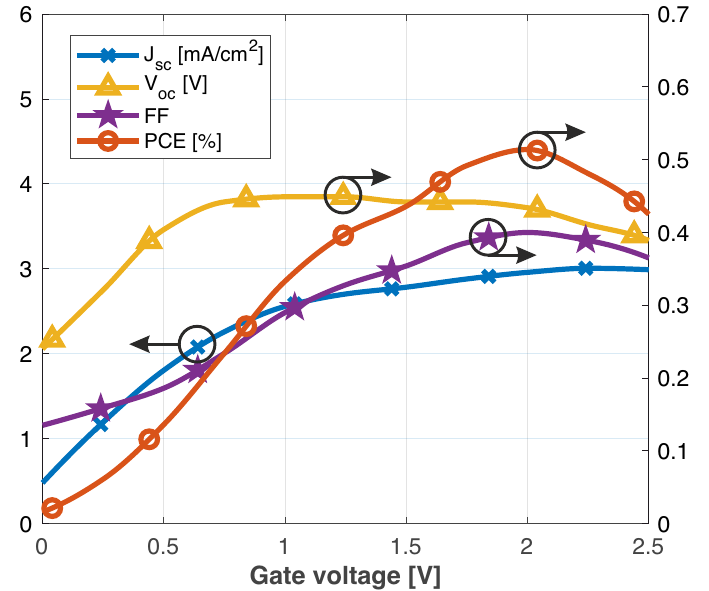}
        \caption{}
    \end{subfigure}%
    ~ 
    \begin{subfigure}[t]{0.325\textwidth}
        \centering
        \includegraphics[width=1\textwidth]{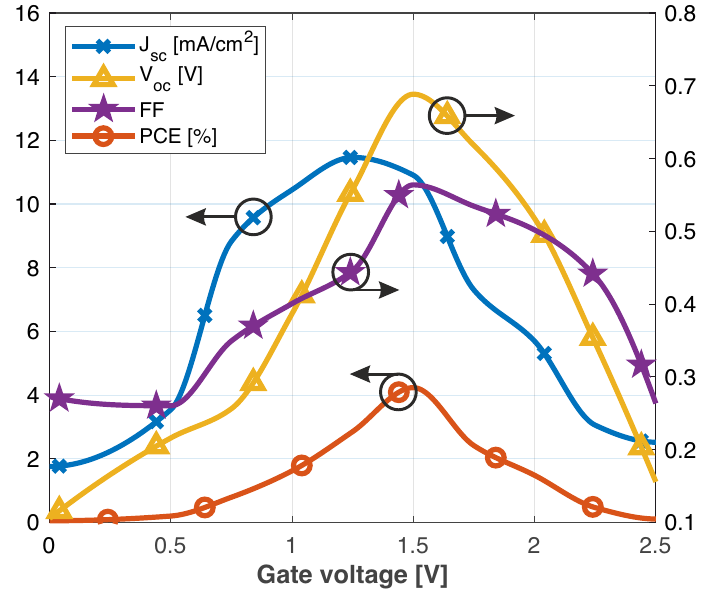}
        \caption{}
    \end{subfigure}%
    ~ 
    \begin{subfigure}[t]{0.325\textwidth}
        \centering
        \includegraphics[width=1\textwidth]{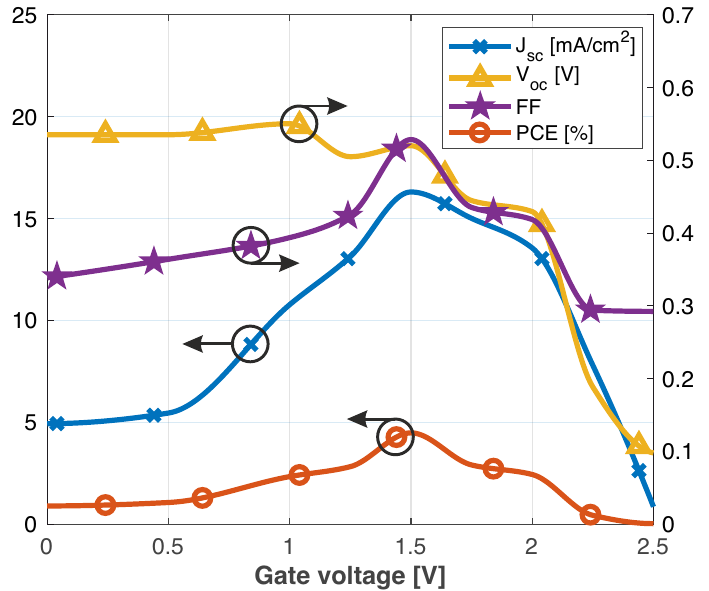}
        \caption{}
    \end{subfigure}
    \caption{Summary of the main extracted parameters: V$\textsubscript{oc}$, FF, J$\textsubscript{sc}$ and PCE change in the following types of OPV: (a) for small molecule OPV with 40 nm ETL thickness, (b) for PTB7:PCBM OPV, and (c) for PTB7:PCBM BHJ with \ce{C60} layer (10 nm) on the top as a function of gate voltage.}
    \label{sm_bhj2}
\end{figure*}

It is quite interesting to compare the observed above dynamic of I-V curves change with similar I-V curves changing upon ionic gating in polymeric BHJ-OPV. Details on PTB7-PCBM BHJ devices are provided in Supplement information (SI). Measured I-V curves and output photovoltaic parameters for polymeric BHJ-OPV are shown below at Figs. \ref{sm_bhj1} and \ref{sm_bhj2}.

The integration of the ion gate to the OPV with PTB7-PCBM bulk heterojunction demonstrated that device operation has several regimes with the increase of V$\textsubscript{g}$ (Fig. \ref{sm_bhj1}a). We found that type of the I-V curve for PTB7:PCBM OPV changed the form from resistor type at initial conditions (no IL at the gate and V$\textsubscript{g}$ = 0 $V$) to diode curve at V$\textsubscript{g}$ = 0.5 -- 1.5 $V$. V$\textsubscript{oc}$ value didn't exceed 0.095 $V$ before contact of CNT cathode with IL. After deposition of DEME-\ce{BF4} V$\textsubscript{oc}$ value increased up to 0.210 $V$. Then, each step (0.25 $V$) with the increase of V$\textsubscript{g}$ showed a large increase of V$\textsubscript{oc}$ up to peak value (0.688 $V$) obtained at V$\textsubscript{g}$ = 1.5 $V$. The gain of J$\textsubscript{sc}$ value with enlargement of the bias at the ion gate corresponds to the improved collection of the charge carriers, the enhanced electric field across the device structure induced by electrostatic accumulation at the CNT/IL interface with the electrical double layer. The absolute increment of J$\textsubscript{sc}$ showed improvement from 0.83 $mA/cm^2$ at initial conditions to 11.47 $mA/cm^2$ at V$\textsubscript{g}$ = 1.25 $V$. As it is shown in Fig. \ref{sm_bhj2}b, the PCE of PTB7:PCBM OPV grew from 0.02 to 4.23\%.

The gain of all output I-V parameters (V$\textsubscript{oc}$, J$\textsubscript{sc}$, FF, PCE) with the increase of V$\textsubscript{g}$ shows that ionic gating turns the operation of the device to a photovoltaic mode that could not be achieved without electrostatic doping with CNT cathode. The clear indicator of this statement is the gain of V$\textsubscript{oc}$ and J$\textsubscript{sc}$, that directly depends on the work function levels of the electrodes forming the built-in field and efficiency of the charge splitting at the BHJ. The FF was improved due to a drop in series resistance (R$\textsubscript{s}$, see SI) and enhancement of the shunt properties at V$\textsubscript{g}$ = 1.0 -- 1.5 $V$. The decrease of the R$\textsubscript{s}$ values was obtained due to the doping of CNT cathode, while an increase of the R$\textsubscript{sh}$ related to the n-type doping of the PCBM layer. However, we should note, that operation of the ionic gate goes with the interaction of interfaces: IL/CNT; IL/PCBM; IL/PTB7. Thus, n-type doping also occurs in the donor layer PTB7 and could harm the efficiency of charge collection.

To exclude the effect of n-type doping at the interface of donor semiconductor, we upgraded the device structure of PTB7:PCBM BHJ with \ce{C60} layer (10 nm) on the top. I-V curve data and output performance for PTB7-PCBM-\ce{C60} device are presented in Figs. \ref{sm_bhj1}b and \ref{sm_bhj2}c, respectively. The J$\textsubscript{sc}$ value was significantly improved in comparison to the OPV structure without \ce{C60} layer. At initial conditions, as well at V$\textsubscript{g}$ values up to 1.50 $V$, J$\textsubscript{sc}$ was higher -- 4.38 $mA/cm^2$ (with \ce{C60}) against 0.83 $mA/cm^2$ (w/o \ce{C60}) at initial conditions, 16.30 $mA/cm^2$ (with \ce{C60}) against 10.49 $mA/cm^2$ (w/o \ce{C60}) at 1.50 $V$. The V$\textsubscript{oc}$ value at initial conditions was much higher for the structure with \ce{C60} --  0.42 $V$, however, at higher voltages at the ion gate V$\textsubscript{oc}$ grew just to 0.52 $V$ due to less favorable level of the LUMO (app. 1 $eV$ higher) in comparison to PCBM. But this point allows making a definite conclusion that resistor behavior of the PTB7-PCBM structure at initial conditions and low V$\textsubscript{oc}$ corresponds to the contact of donor material with the electrode and n-type doped area near the electrical double layer formed by the IL.

Comparing PV parameters depending on gate voltage in Fig. \ref{sm_bhj2} we may note that curves demonstrate quite different qualitative behavior: in BHJ-OPV case (b) the main effect is the dramatic change of V$\textsubscript{oc}$ upon V$\textsubscript{g}$ while in small molecule device (a) V$\textsubscript{oc}$ is not increased significantly and it is already high enough even at V$\textsubscript{g}$=0 and absence of IL, although the Fermi level is known to raise in CNTs. This insensitivity of V$\textsubscript{oc}$ on ionic gate V$\textsubscript{g}$ indicates that in planar small molecule with ETL the LUMO of C$\textsubscript{60}$ acceptor determines V$\textsubscript{oc}$, and the work function of CNT is not that important. 
This different physical behavior refers to different role of interfaces in collecting J$\textsubscript{sc}$. In our small molecule OPV the largest effects are an increase of J$\textsubscript{sc}$ and qualitative change of S-shaped FF into box type FF by better collection of charge carriers and partially by significant lowering the series R$\textsubscript{s}$ due to dramatically changes of CNT sheet resistance.

Another difference in I-V behavior of the PTB7-PCBM based OPV comparing to small molecule based device is the appearance of the drop of the performance after peak value (1.5 $V$ for PTB7-PCBM based OPV), while the operation of small molecule based device at V$\textsubscript{g}$ > 1.5 $V$ still shows reasonable values of the PCE (Fig. \ref{sm_bhj2}a). This effect is related to the electrochemical oxidation process with a reaction between DEME ions and polymer bones of PTB7, as it was shown for other polymers under ionic gating conditions. Thicker \ce{C60} layer protects CuPc from the ions of DEME allowing higher V$\textsubscript{g}$. For instance, small molecule device with a 5 nm thick ETL reaches its best performance at lower V$\textsubscript{g}$ = 0.85 $V$ as shown in Fig. \ref{C60_5nm_all}b.


\subsection{Numerical simulations by drift-diffusion model}

To support our interpretation of the experimental results, we performed a 1-D numerical analysis of small molecule OPV system by coupling of optical (transfer-matrix formalism) and electrical (drift-diffusion model) simulations. To model free charge carrier generation inside the active layer, we used the Onsager-Braun theory and considered Langevin recombination as the dominant process for bound charge recombination in our organic solar cell, as described below.

For the optical simulation, we applied Transfer-Matrix Method thereby taking into account the interference phenomena in the whole multilayered structure and used the AM1.5G standard spectrum as the light source to find the electric field distribution inside the active region. On the other hand, the simulation domain in the electrical part was restricted by the electrodes with appropriate boundary conditions.

The main process that takes place in OPV is the generation of free charge carriers due to the absorption of a photon in the active region. The absorbed photon creates an exciton which can freely move until it dissociates into free charge carriers at a donor/acceptor interface. In this model, we did not take into account excitons dynamic and imagine that they instantly dissociated to polaron with 100\% probability defining the lifetime only for polaron. Therefore instead of considering a morphology of the bulk heterojunction we replaced a blend layer with an effective medium and used the drift-diffusion model. The advantage of the drift-diffusion model is that it can predict the overall behavior of OPV effectively at the price of a much less computational complexity comparing with other approaches \cite{fallahpour2014optoelectronic, koh2011three, koster2005device}. 

\subsubsection{System of equations}
In the drift-diffusion model, the basic equations that should be solved are Poisson's equation:

\begin{equation} \label{poisson}
\nabla^2 V=\rho/\epsilon,
\end{equation}
where \(\rho \) is the density of charge, \(V \) is the potential and \(\epsilon \) is the relative permittivity, and the continuity equation:

\begin{equation} \label{continuity_eqn}
\nabla \cdot J_i=\pm qU   \,\,\,\,\,\,\,\,\,\,\,    i=n,p,
\end{equation}
where \(J_i\) is the current density for electron and hole, \(q\) is the elementary charge and \(U\) is the net generation rate of charge carriers.
In order to complete our system of equation we should take into account drift-diffusion equation for electrons and holes:
 
\begin{equation} \label{dd1_eqn}
J_n=qn\mu_n\nabla V\textsubscript{LUMO} + \mu_n k_BT\nabla n
\end{equation}
\begin{equation} \label{dd2_eqn}
V\textsubscript{LUMO} = - (V+\xi)
\end{equation}
\begin{equation} \label{dd3_eqn}
J_p=qn\mu_p\nabla V\textsubscript{HOMO} + \mu_p k_BT\nabla p
\end{equation}
\begin{equation} \label{dd4_eqn}
V\textsubscript{HOMO} = - (V+ E\textsubscript{gap} + \xi),
\end{equation}
where the first terms in Eqs. (\ref{dd1_eqn}) and (\ref{dd3_eqn}) are responsible for the drift of charges due to the electric field and the second terms are for the diffusion of particles due to the gradient of their concentration. Here \(k_B\) is the Boltzmann constant, \(T\) is the absolute temperature, \(\mu_n\) and \(\mu_p\) are the mobilities of electrons and holes, respectively, \(n\) and \(p\) are the concentration of carriers for electrons and holes, \(\xi\) is the electron affinity \(V\textsubscript{LUMO}\) is the energy level of LUMO band, \(V\textsubscript{HOMO}\) is the energy level of HOMO band and \(E\textsubscript{gap}\) is the energy difference between LUMO and HOMO levels. We assume that the initial concentration of carriers obeys the Fermi-Dirac distribution. 

\subsubsection{Boundary conditions}

In our simulations, each electrode of the OPV imposes boundary conditions. The contact properties are defined by the energy offset between the Fermi energy of the metal and HOMO and LUMO levels of the semiconductor. According to the thermionic emission theory, charge carriers can injected from the metal into the semiconductor's bands by thermal activation. Therefore, for a cathode with injection barrier \(\phi_n\) (the potential difference between the work function of cathode and LUMO level of semiconductor), injected charge carrier densities are given by:

\begin{equation} \label{metal_contact}
n\textsubscript{th}=N\textsubscript{LUMO} \cdot exp \bigg( \frac{-q\phi_n}{k_BT} \bigg)
\end{equation}
\begin{equation} \label{metal_contact2}
p\textsubscript{th}=N\textsubscript{HOMO} \cdot exp \bigg( \frac{-q(\phi_n-E\textsubscript{gap})}{k_BT} \bigg),
\end{equation}
where \(N\textsubscript{LUMO}\) and \(N\textsubscript{HOMO}\) are the effective density of states for LUMO and HOMO bands, respectively. For anode, it can be described accordingly:

\begin{equation} \label{eu_eqn}
n\textsubscript{th}=N\textsubscript{LUMO} \cdot exp \bigg( \frac{-q(\phi_p-E\textsubscript{gap})}{k_BT} \bigg)
\end{equation}
\begin{equation} \label{eu_eqn}
p\textsubscript{th}=N\textsubscript{HOMO} \cdot exp \bigg( \frac{-q\phi_p}{k_BT} \bigg),
\end{equation}
where \(\phi_p\) is the barrier between HOMO level and Fermi level of anode.

Moreover, the recombination of charge carriers takes place at surfaces of electrodes \cite{grove}. Two charge carriers with opposite signs recombine if both enter simultaneously in metal electrode. The surface recombination currents of excess charges are defined as $J\textsubscript{(s,n)}=qs_n(n-n\textsubscript{th})$ for electrons and $J\textsubscript{(s,p)}=qs_p(p-p\textsubscript{th})$ for holes. Here, the surface recombination velocities \(s_n\), \(s_p\) denote the speed and, consequently, the efficiency of the recombination process. The boundary condition for the potential reads $V\textsubscript{anode}-V\textsubscript{cathode}=V_a-V\textsubscript{bi}$, where \(V_a\) is the applied voltage and  \(V\textsubscript{bi}\) is built-in voltage that is obtained from difference between the work function of electrodes. 

\subsubsection{Free charge carrier generation}
The charge carriers can be created via the dissociation of excitons in blend and this is described by Onsagers theory for electrolyte dissociation. According to the theory, polaron pairs with binding distance \(r\) under an applied electric field \(\nabla V\) are separated with the rate \(k\textsubscript{diss}\):

\begin{equation} \label{eu_eqn}
k\textsubscript{diss}=\frac{3\gamma}{4\pi a^3} exp \bigg(\frac{E_b}{k_B T} \bigg)\frac{J_1(2-\sqrt{-2b})}{\sqrt{-2b}},
\end{equation}
where \(J_1\) is the first order Bessel function with its argument $b=q^3\nabla V / (8\pi\epsilon_0\epsilon_r k_B^2T^2)$, \(E_b\) is the Coulomb binding energy between two charged particles, \(\gamma\) is the Langevin recombination factor: $\gamma=q (\mu_n+\mu_p) / (\epsilon_r\epsilon_0)$. For solving this equation numerically, we can use the expansion of Bessel function:

\begin{equation} \label{eu_eqn}
k\textsubscript{diss}=\frac{3\gamma}{4\pi a^3} exp \bigg(\frac{E_b}{k_B T} \bigg)(1+b+\frac{b^2}{3}+\frac{b^3}{18}+\frac{b^4}{180}+...).
\end{equation}

The temperature dependent polaron pair separation \(k\textsubscript{diss}(r,T,\nabla V)\) competes polaron pair decay rate to the ground state \(k_F(T)\). Therefore, separation probability of a polaron pair with a binding distance \(r\) under an external electric field \(\nabla V\) is defined by:
 
\begin{equation} \label{probability}
p=\frac{k\textsubscript{diss}}{k\textsubscript{diss}+k_F}.
\end{equation}

The dissociation rate is integrated over a Gaussian distribution of separation distances to considered local disorder in organic semiconductor material. Considering the most probable binding distance \(a\), the integral over all binding distances \(r\) results in an average polaron pair separation probability \(P\):
\begin{equation} \label{eu_eqn}
P=\int_{0}^{\infty} p\frac{4}{\sqrt{\pi}a^3}r^2 exp \bigg(\frac{-r^2}{a^2} \bigg) dr.
\end{equation}

 Since the donor-acceptor interface is distributed throughout the volume of BHJ, we assume that generated exciton in blend is converted to polaron with 100\% probability. Therefore, then the number of polaron per unit volume \(X\) is changed in time by: \cite{barker2003modeling}
\begin{equation} \label{eu_npolaron}
\frac{dX}{dt}=G-k_F X-k\textsubscript{diss}X+R,
\end{equation}
where \(G\) is the generation rate of polaron, \(R\) is the recombination rate.

In a steady state case ($dX/dt=0$), we can express the continuity equation for electrons from Eqs. (\ref{continuity_eqn}), (\ref{probability}) and (\ref{eu_npolaron}) as following:

\begin{equation} \label{net_G}
\nabla \cdot J_n=-q(PG-(1-P)R),
\end{equation}
where the right side of Eq. (\ref{net_G}) is the net generation rate.

\subsubsection{Recombination process}
The dominate charge carrier recombination process in organic semiconductors is the Langevin recombination \cite{langevin1903recombinaison} which is defined by: 
\begin{equation} \label{eu_rec1}
R_L=\zeta\gamma(np-n_i^2).
\end{equation}
 Here \(\zeta\) is the correction coefficient that applied to the Langevin recombination factor because in experiment the recombination rates are different than what predicted by the Langevin theory \cite{mingebach2012dirac}. In Eq. (\ref{eu_rec1}), \(n_i\) is the intrinsic charge carrier defined as following: $n_i=\sqrt{N\textsubscript{HOMO}N\textsubscript{LUMO}} \cdot exp ( -E\textsubscript{gap} / (2k_B T))$.

\subsubsection{Exciton generation}

For calculation of exciton generation rate, we should find the absorbed energy of light inside the active layer which is given by:
\begin{equation} \label{eu_eqn}
Q(x,\nu)=\frac{1}{2}c\epsilon_0\alpha n \mid E(x)\mid^2,
\end{equation}
 where \(c\) and \(\epsilon_0\) are the speed of light in the vacuum and the permittivity of free space, respectively, \(\alpha\)  and \(n\) are the absorption coefficient and the refractive index of the active layer, and \(E\) is the electric field intensity. The generation for each frequency at each point is determined by $G(x,\nu)=Q(x,\nu) / (h\nu)$. Integration of $G(x,\nu)$ over all the frequencies allows us to find the exciton generation density as a function of the coordinate $G(x)=\int_{}^{} G(x,\nu)d\nu$.

\subsubsection{Fitting the observed I-V data}

 \begin{table}
  \caption{Materials parameters used in electrical simulations. \cite{su2007work,tokunaga,Berkai2017,chen2006}}
  \label{tb2_parameters}
  \begin{tabular}{llllll}
    \hline
    \hline
    Parameter	&	Symbol	&	HTL &	CuPc:C$_{60}$	&	C$_{60}$	&	Unit\\
    \hline
	Temperature	&	T	&	300	&	300	&	300	& \(K\)	\\
	LUMO Energy	&	E\textsubscript{LUMO}	&	-3.90	&	-4.00	&	-4.12   &   \(eV\) \\
	HOMO Energy	&	E\textsubscript{HOMO}	&	-4.95	&	-5.80	&	-6.42	& \(eV\)	 \\
	Layer Thickness	&	L	&	7	&	60	&	40	& \(nm\) \\
	Electron Mobility	&	\(\mu_n\)	&	9.04e-4	&	2.0e-3	&	6.0e-3	&  \(cm^2/(V\cdot s)\)	\\
	Hole Mobility	& \(\mu_p\)	&	1.0e-2	&	4.0e-4	&	2.0e-5	&	\(cm^2/(V \cdot s)\) \\
	Effective Density of States	&	\(N\textsubscript{LUMO}, N\textsubscript{HOMO}\)	&	1.3e22	&	1.3e22	&	1.3e22	&	\(1/m^3\)	\\
	Binding Distance*	&	a	&	1.34	&	1.34	&	1.34	&	\(nm\) \\
	Langevin coefficient Correction*	&	\(\zeta\)	&	0.8	&	0.8	&	0.8	&  -- \\
	Decay Rate	&	\(k_F\)	&	8e5	&	8e5	&	8e5	& \(1/s\)	 \\
	Relative Permittivity	&	\(\epsilon_r\)	&	3.4	&	3.4	&	4 & --	\\
	ITO work function	&	\(WF\textsubscript{ITO}\)	&	-4.4	&	--	&	--	&	\(eV\)	\\
	MWCNT work function	&	\(WF\textsubscript{MWCNT}\)	& --		& -- &	-4.76	&	\(eV\)	\\
	Surface Recombination Velocity 	&	\(S_n,S_p\)	&	2e5	&	--	& --	& \(m/s\)	\\
    \hline
    \hline
  \end{tabular}
    \begin{tablenotes}
        \item[a] *This parameter used as fitting factors for the reconciling experimental and numerical data.
    \end{tablenotes}  
\end{table}

For considering the IL gating process in our system, we used indirect estimation techniques based on fitting experimental I-V curves. We assumed that the ions that penetrate inside MWCNTs reduce the work function of the cathode. Increasing gate voltage corresponds to higher penetration of ions and consequently rising of the cathode Fermi level. Moreover, penetration of ions is not limited to the cathode region -- ions can also move further inside the ETL layer through MWCNT porous media. This process is shown in Fig. \ref{IL_penetration} and implemented as external n-type doping in our simulations. The material parameters that we used in electrical simulations are summarized in Table~\ref{tb2_parameters}.

\begin{figure}[htbp]
\centering
\includegraphics[width=0.7\textwidth]{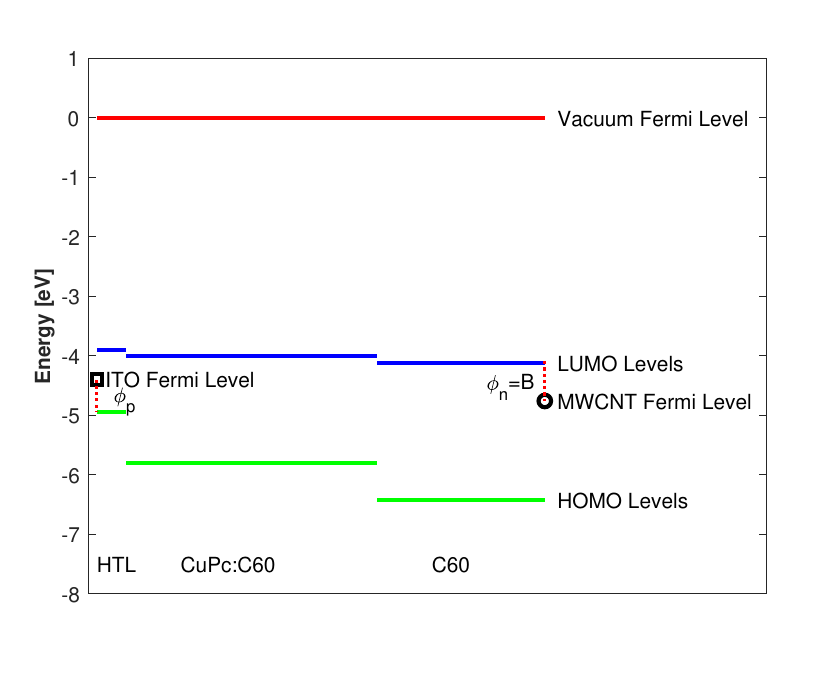}
\caption{Energy diagram of our OPV system. Here \(B\) denotes the potential barrier that electrons experience during transition from LUMO level of \ce{C60} to MWCNT cathode.}
\label{Energy_level}
\end{figure}

\begin{figure}[htbp]
    \centering
    \begin{subfigure}[t]{0.475\textwidth}
        \centering
        \includegraphics[width=1\textwidth]{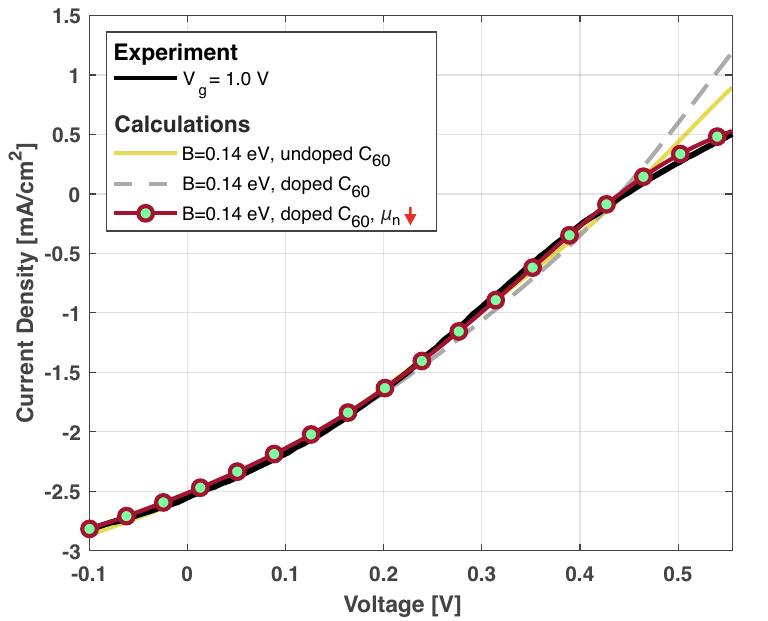}
        \caption{}
    \end{subfigure}%
    ~ 
    \begin{subfigure}[t]{0.455\textwidth}
        \centering
        \includegraphics[width=1\textwidth]{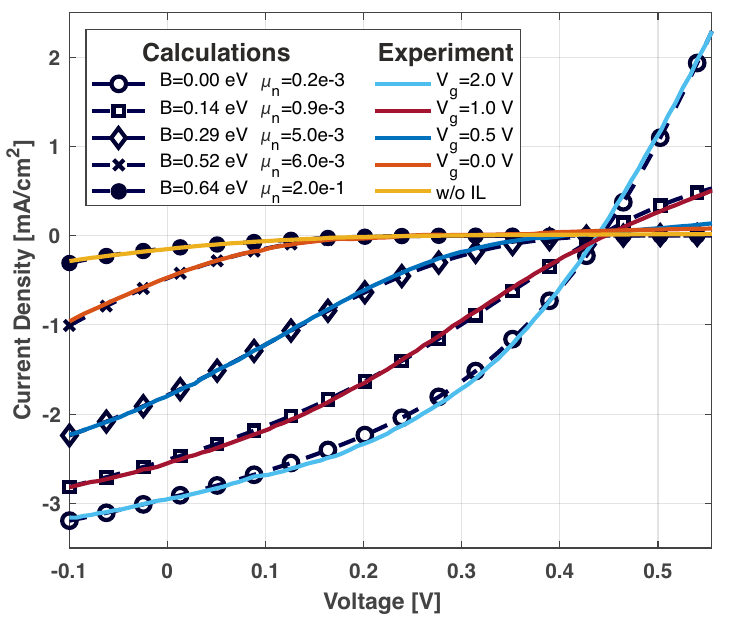}
        \caption{}
    \end{subfigure}
    \caption{Comparison of numerical and experimental I-V curves for \ce{C60}-based device with 40 nm thick ETL considering \ce{C60} doping and electron mobility variations (mobility unit is \(cm^2/(V \cdot s\) ))  as a function of gate voltage (b) with a specific example of fitting procedure for V$\textsubscript{g} = 1 V$ (a).}
    \label{IV_BDM}
\end{figure}

In Fig. \ref{Energy_level}, the maximum energy difference between LUMO level of \ce{C60} and MWCNT work function is 0.64 $eV$ that appeared as \(\phi_n\) in Eqs. (\ref{metal_contact}) and (\ref{metal_contact2}). We next denote it as a barrier in the text. At first, we suppose that ions penetrate only inside MWCNTs and this penetration only reduces the work function of the cathode and hence results in a reduction of the barrier. We chose 0 \(V\) gate voltage case and fitted the numerical I-V curves to experimental ones through the selection of proper fitting factor and barrier value. Then, we fitted I-V curves for other gate voltages by only changing of barrier value. Although this approach gives a relatively good match for fitted data it doesn't allow to reproduce exact shapes of experimental I-V curves. The most noticeable difference between numerical data and experimental results was observed for gate voltages above 1 \(V\) as shown in Fig. \ref{IV_BDM}a with yellow and black solid curves, respectively (we don't provide curves for other gate voltages to avoid cluttering at this intermediate step).

At this point, we assume that the penetration of positive ions from ionic liquid goes beyond of MWCNT and reach to the \ce{C60} layer, and because of the existance of tetrahedral and octahedral interstitial voids between \ce{C60} spheres of large enough volume to accommodate ions of IL the positive ions penetrate into these voids by Faradaic process of \ce{C60} redox and charging \ce{C60} to \ce{C60} n-type ion, where the number of electrons in \ce{C60} depend on the \(V_g\) and concentration of ions in voids. We suppose that penetration depth and amplitude are proportional to gate voltage and with increasing gate voltage (reduction of a barrier) more ions penetrate inside \ce{C60} layer. The injected ion to \ce{C60} layer produces extra free charge. This free charge contributes to charge density that appeared in right-hand side of Eq. (\ref{poisson}). Moreover, penetrated positive ions can affect the mobility of electron in \ce{C60} layer. Therefore, we first used binding distance \(a\), Langevin coefficient correction \(\zeta\), barrier \(\phi_n\) and electron mobility \(\mu_n\) parameters for fitting at 0\(V\) gate voltage. Then, we used only barrier and electron mobility as the fitting parameters for other gate voltages. New parameter value that derived for fitting of 0V bias voltage was 1.32 $nm$ for \(a\) and \(1\) for \(\zeta\).

Comparing curves in Fig. \ref{IV_BDM}a one can see that considering penetrated ion into \ce{C60} layer and the effect of this penetration on electron mobility in \ce{C60} layer in the model allows us to fit I-V curve more accurately. In Fig. \ref{IV_BDM}b we show the fitted I-V curves for different gate voltages. From this figure, one can see that for 0V gate voltage the obtained barrier is less than the maximum barrier. This means that even without gate voltage applied the IL can change the barrier slightly. Also, the electron mobility in the case of the absence of the IL is much higher than that for the case of the presence of the IL. This difference is two orders of magnitude, therefore we can conclude that injection of the IL inside \ce{C60} layer drops the electron mobility drastically. Although the barrier value derived from the fitting of the case without the IL is larger than the maximum barrier, this difference is quite small and negligible.

\begin{figure*}[t!]
    \centering
    \begin{subfigure}[t]{0.45\textwidth}
        \centering
        \includegraphics[width=1\textwidth]{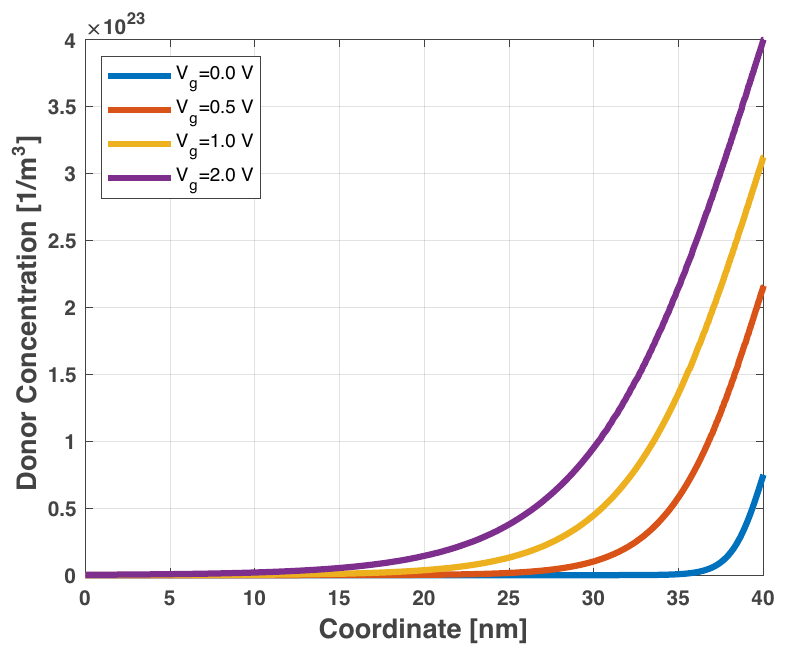}
        \caption{}
    \end{subfigure}%
    ~ 
    \begin{subfigure}[t]{0.455\textwidth}
        \centering
        \includegraphics[width=1\textwidth]{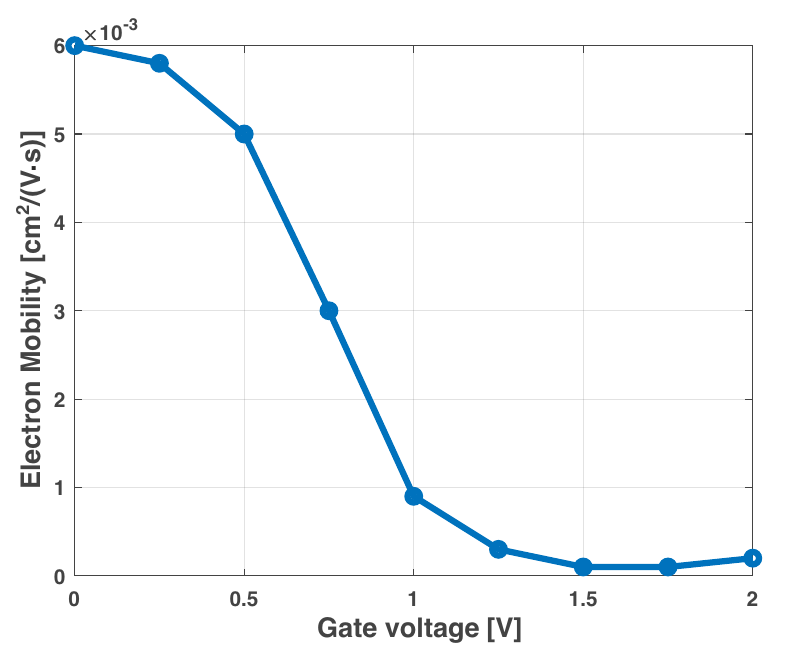}
        \caption{}
    \end{subfigure}
    \caption{Donor concentration (a) -- meaning positive ions from IL that diffused into \ce{C60} upon V$\textsubscript{g}$ exceeding the electrochemical potenitial of \ce{C60} (i.e. at V$\textsubscript{g}$ becoming higher than bottom of \ce{C60} conduction band) and electron mobility (b) in \ce{C60} layer as a function of penetration depth and gate voltage, respectively. }
    \label{Donor-mobility}
\end{figure*}

Fig. \ref{Donor-mobility} shows the donor concentration profile in \ce{C60} layer and the derived value for mobility from the fitting of I-V curve with respect to the gate voltage. In this figure one can see that for small gate voltage that corresponds to low ion penetration inside \ce{C60} layer, mobility of electron is high and with increasing of ion penetration the mobility of electron decreases until it reaches a constant value. This constant value corresponds to the case when the IL saturate the \ce{C60} layer.

\begin{figure*}[t!]
    \centering
    \begin{subfigure}[t]{0.45\textwidth}
        \centering
        \includegraphics[width=1\textwidth]{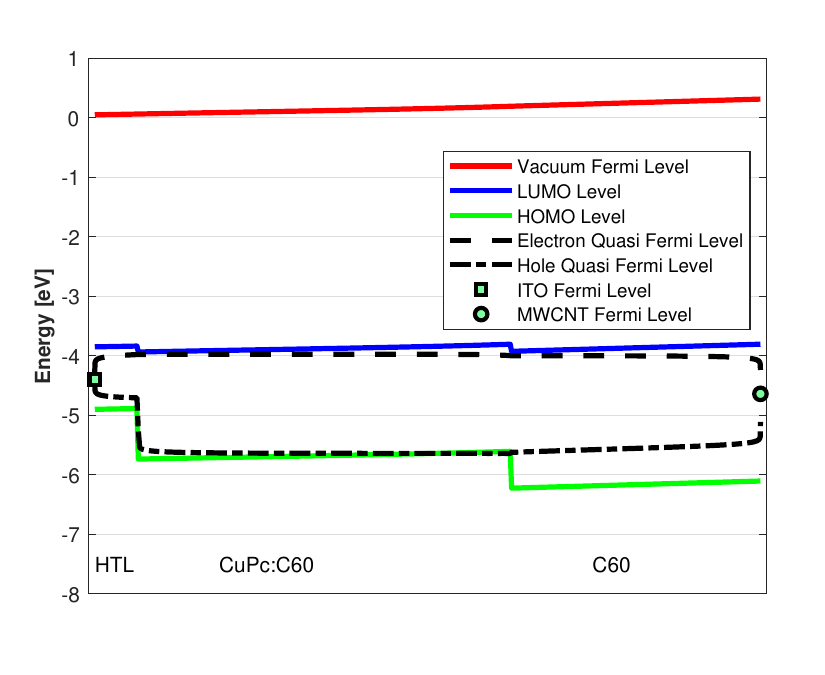}
        \caption{}
    \end{subfigure}%
    ~ 
    \begin{subfigure}[t]{0.45\textwidth}
        \centering
        \includegraphics[width=1\textwidth]{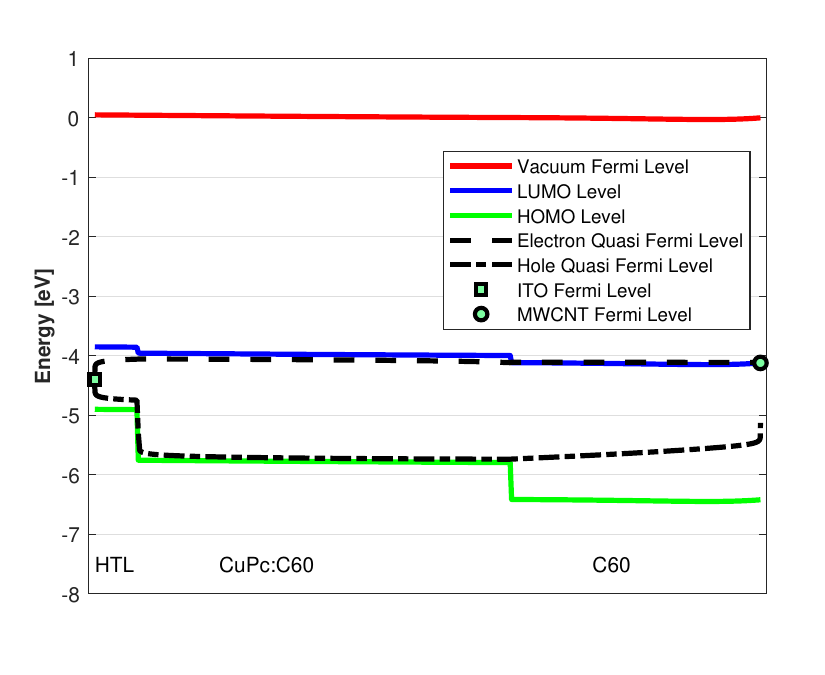}
        \caption{}
    \end{subfigure}
    \caption{Energy diagram of our OPV-IL system for 0 \(V\) (a) and 2 \(V\) (b) gate voltage at the maximum extraction power point corresponding to bias voltage 0.075 \(V\) and 0.28 \(V\), respectively}
    \label{E_band}
\end{figure*}

Fig. \ref{E_band} shows the energy diagram of OPV for different gate voltages. In Fig \ref{E_band}b gate voltage is 2 \(V\), therefore barrier is zero and more ion can penetrate inside \ce{C60} layer. From this figure, one can see the bending of energy level for \ce{C60} layer at the contact point with MWCNT. This bending is formed because of presence of the IL inside \ce{C60} layer. Moreover, the comparison of Fig. \ref{E_band}a and \ref{E_band}b gives us an understanding of how changing of cathode Fermi level fits the quasi-Fermi level of electron to cathode work function and that is the reason of considerable enhancement of Fill factor for small barrier.

\section{Conclusions}

We have demonstrated a tunable process by ionic gating small molecule OPV structure with up to 200 nm-thick stable fullerenes (\ce{C60} and \ce{C70}) ETL with a highly porous MWCNT n-dopable charge collector with tuned Fermi level by EDLC charging. This structure shows an improvement of all PV parameters V$\textsubscript{oc}$, J$\textsubscript{sc}$ and FF leading to more than 50-time increase in PCE upon optimal ionic gating. Such operation is usually achieved with a very thin intrinsic fullerene layer of 7 to 10 nm. Our numerical results show a good consistency of I-V curve behavior and the main photovoltaic parameters with the experimental data. Our modeling allows to understand the difference of dynamics of I-V curves change during ionic gating in small molecule OPV as compared to polymeric BHJ-OPV, that was observed earlier~\cite{cook2014ambient, cook2013electrochemically, kim2012semi, saranin2017tunable}. This difference is explained by differences in geometry of interfaces, i.e. mostly planar interface of small molecule OPV and geometry of small molecule vs BHJ extended interpenetrating network and it can be further analyzed quantitatively by suggested here drift-diffusion model.

\begin{acknowledgement}

This work was partially supported by the Ministry of Education and Science of the Russian Federation (Grant 14.Y26.31.0010).

\end{acknowledgement}

\bibliography{opv_ionic}

\end{document}